\documentclass[useAMS,usenatbib]{biom}
\usepackage{hyperref} 
\hypersetup{colorlinks, allcolors = blue, citecolor=blue, urlcolor=black}

\usepackage[dvipsnames]{xcolor} 
\usepackage{amsmath, amsfonts, bbm}
\usepackage{booktabs}
\usepackage{mathtools}
\usepackage{pdfpages} 

\def\bSig\mathbf{\Sigma}

\newcommand{\indep}{\perp \!\!\! \perp}

\title[A Causal Inference Framework for Leveraging External Controls in Hybrid Trials]{A Causal Inference Framework for Leveraging External Controls in Hybrid Trials}

\author{Michael Valancius$^{1,*}$\email{mval@email.unc.edu}, Herbert Pang$^{2}$,  \textbf{Jiawen Zhu}$^{2}$, \\
\textbf{Stephen R Cole}$^{3}$, \textbf{Michele Jonsson Funk}$^{3}$, \textbf{and Michael R Kosorok}$^{1}$ \\
$^{1}$Department of Biostatistics, University of North Carolina at Chapel Hill, Chapel Hill, NC, U.S.A. \\
$^{2}$Product Development Data Sciences, Genentech, South San Francisco, CA, U.S.A. \\
$^{3}$Department of Epidemiology, University of North Carolina at Chapel Hill, Chapel Hill, NC, U.S.A. }

\begin{document}







\label{firstpage}


\begin{abstract}
We consider the challenges associated with causal inference in settings where data from a randomized trial is augmented with control data from an external source to improve efficiency in estimating the average treatment effect (ATE). Through the development of a formal causal inference framework, we outline sufficient causal assumptions about the exchangeability between the internal and external controls to identify the ATE and establish the connection to a novel graphical criteria. We propose estimators, review efficiency bounds, develop an approach for efficient doubly-robust estimation even when unknown nuisance models are estimated with flexible machine learning methods, and demonstrate finite-sample performance through a simulation study. To illustrate the ideas and methods, we apply the framework to a trial investigating the effect of risdisplam on motor function in patients with spinal muscular atrophy for which there exists an external set of control patients from a previous trial. \\
\end{abstract}

%


\maketitle


%

\section{Introduction} \label{sec_intro}

Establishing the causal effect of a novel intervention is imperative for making informed decisions about its adoption or approval. Randomized clinical trials (RCTs) generate robust causal evidence by ensuring treatment assignment is independent of baseline factors. However, some situations warrant alternative study designs. Investigators might have access to additional patient records under the control intervention \citep{yue_leveraging_2018}, while rare-disease settings face ethical or feasibility concerns with enrolling patients into standard RCTs \citep{jansen-van_der_weide_rare_2018, massicotte_open-label_2003}.  As a motivating example, we consider a study of risdiplam's effect on motor functioning for patients with spinal muscular atrophy (SMA) \citep{mercuri_safety_2022}. While the trial's control arm is small, patient data exists from a historical control arm. More generally, we consider hybrid trials where study participants are randomized to treatment via a known mechanism and externally-collected control patient records (external controls) are available for analysis \citep{baumfeld_andre_trial_2020}. Amidst heightening emphasis on utilizing real-world data \citep{corrigan-curay_real-world_2018} and increasing regulatory approval of hybrid trials \citep{hatswell_regulatory_2016}, we investigate the implications of adopting these trials on our ability to estimate causal effects. The analysis draws focus to two primary themes.  \textbf{(1)} Randomization to treatment is no longer (fully) under investigator control. \textbf{(2)}  Adjustment for imbalances due to a lack of randomization can increase dependence on statistical models.

Our major contributions are formally characterizing the setting's structure and providing an analysis framework for estimating average treatment effects. To address (1), we present a causal inference framework for rigorously defining the target parameter, detailing assumptions, and exploring potential sources of bias. Leveraging advances in statistical causal inference theory \citep{kennedy_semiparametric_2023,chernozhukov_doubledebiased_2018}, we alleviate (2) by extending efficient estimators of the average treatment effect (ATE) to allow for estimating nuisance functions with machine learning methods and establishing theory for valid statistical inference.
\subsection{Related Work}
The challenges associated with using external controls in favor of a traditional RCT have been studied for at least half of a century \citep{pocock_combination_1976, sacks_sensitivity_1983, fleming_historical_1982}, with Pocock's 1976 criteria becoming a frequently-referenced standard for evaluating the suitability of external controls. More recently, others have outlined practical considerations and potential sources of bias in hybrid trials with external controls \citep{ghadessi_roadmap_2020, schmidli_beyond_2020, hall_historical_2021}.

Our work expands upon and provides a justification for causal inference methods for trials with external controls. The closest work, which forms an important basis for our discussion on estimation, uses semiparametric efficiency theory to derive the efficient influence curve of the ATE, leading to a doubly-robust estimator \citep{li_improving_2021}. Propensity score methods estimate the conditional probability of an individual receiving the treatment and then use this estimated probability to either match external controls to study participants \citep{lin_propensity_2018} or inversely weight external controls \citep{magaret_new_2022}, Others have focused on estimating the conditional average treatment effect (CATE) \citep{zhou_incorporating_2021}. A related topic is the more general problem of causal inference when combining data from multiple sources \citep{dunipace_optimal_2022, shi_data_2023}. An alternative approach to incorporating external controls forgoes an explicit causal model, borrowing information from external controls through an adaptive Bayesian prior. The amount of borrowing adapts to heterogeneity between the internal and external controls, with external controls whose outcome conflicts with internal controls contributing less to the prior \citep{schmidli_robust_2014, ibrahim_power_2015, liu_propensity-score-based_2021}

\subsection{Outline}

In this work, we develop a framework for the causal analysis of a randomized trial incorporating external controls. We define the target parameter, outline sufficient causal assumptions to link the parameter to the external control data, and propose a graphical method for to help evaluate the credibility of these assumptions. We outline three general strategies for estimating the target parameter and demonstrate their dependence upon unknown nuisance functions. Based upon efficiency theory, we build upon the previously proposed doubly-robust estimator \citep{li_improving_2021} and prove conditions under which the estimator is efficient and asymptotically normal even when machine learning methods are used for the nuisance functions, an important advance that allows for valid inference under broader data generating mechanisms. The variance reduction and double-robustness is explored through a simulation study, and the entire causal framework is demonstrated through a study of patients with spinal muscular atrophy.

\section{Setting}
\label{s:setting}

\subsection{The Target Trial}

In a hybrid trial design with external controls, data from two sources are used to estimate a treatment effect \citep{baumfeld_andre_trial_2020}. Precisely defining this objective requires carefully answering two questions. \textbf{(1)} What two populations generated the data? Differences between the populations generating the data governs both the capacity to combine the two data sources and the methods to do this accurately and efficiently. \textbf{(2)} What population's treatment effect? Unless the treatment effect is constant, the ATE varies across populations.

\begin{figure*}
    \centering
\includegraphics[width = 7.0in]{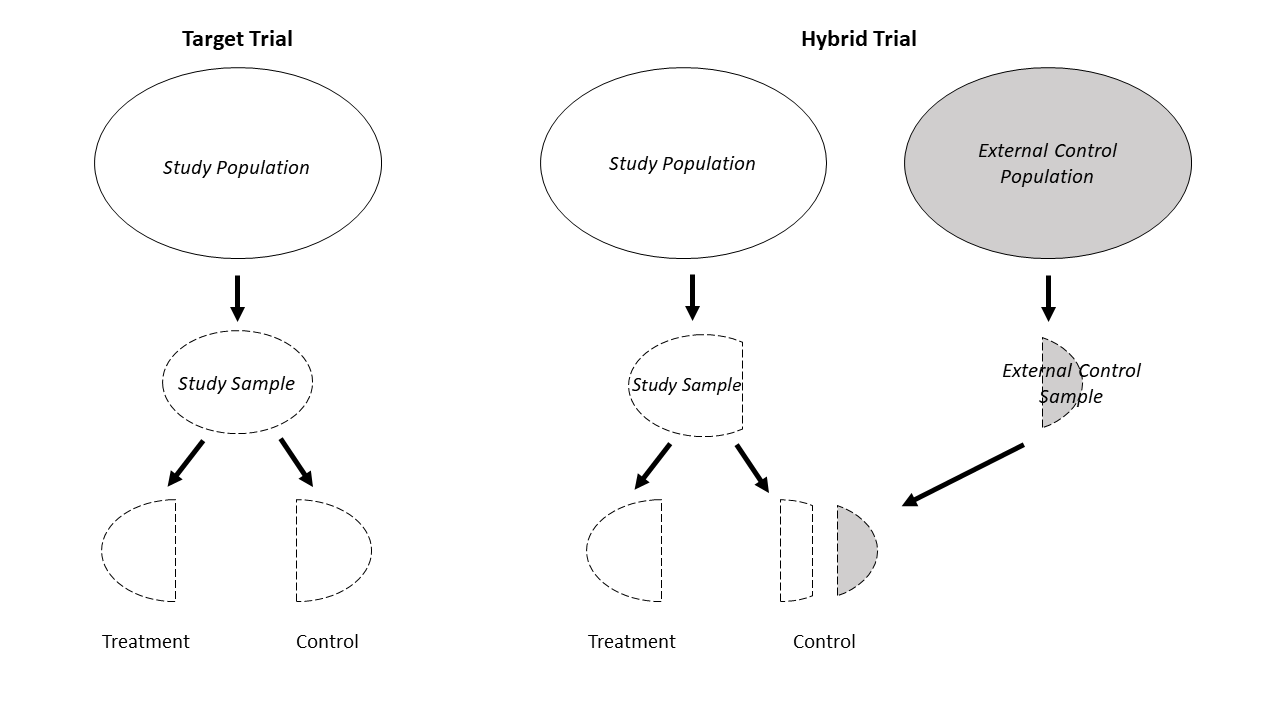}
\caption{A comparison of the target trial (left) and the hybrid trial with external controls (right). Elements from the study population and external control population are white and gray, respectively.}
\label{fig:study_design}
\end{figure*}

Because randomization no longer fully describes treatment assignment, these studies share challenges similar to those of observational studies. For this reason, we recommend evaluating the study through the lens of a target trial  \citep{hernan_causal_2023}. Briefly, a target trial is the hypothetical (fully) randomized trial that a non-randomized study would like to emulate. Our target trial is a fully randomized study sampling $n$ participants from the study population, where $n$ is the number of internal and external records in the hybrid trial (Figure \ref{fig:study_design}).  Adopting this perspective is useful for at least two reasons. First, it orients the goal of the analysis as having, at a minimum, high internal validity. Second, it forces researchers to critically assess the suitability of a proposed external control arm in the design stages. For example, can the same inclusion/exclusion criteria be applied to the external control arm? Can the treatment strategy of the external control arm reasonably resemble the same treatment strategy of the internal control arm, including the degree of monitoring and enforcement of adherence? If the answer to these questions is "no," then it might be sensible to preemptively modify the randomized portion of the trial (and thus the target trial) or find a more compatible external control cohort.

\subsection{Data Structure}

Definitions of populations relevant to inference vary across contexts and discipline; we adopt definitions commonly used within the causal inference literature \citep{laan_targeted_2011}. The \textit{study sample} consists of (internal) enrolled study participants. The \textit{study population} is the (possibly hypothetical) population from which the study sample is a simple random sample. The \textit{target population} is the population of investigator interest, but is rarely the same as the study population due to logistical barriers, scientific considerations, or ethical concerns \citep{rothwell_factors_2006, bell_estimates_2016}. Differences in treatment effects between the study and target populations leads to a lack of \textit{external validity}; combating this challenge is the central focus of the research disciplines of generalizability and transportability \citep{cole_generalizing_2010, bareinboim_general_2013, lesko_generalizing_2017, westreich_transportability_2017, dahabreh_generalizing_2019}. Investigators also have access to data from a second sample of individuals, the external controls. These external controls might have come from one of several sources, including previous clinical trials \citep{darras_risdiplam-treated_2021} and electronic health record (EHR) systems \citep{carrigan_using_2020}, and it is implicitly understood that they have been selected to satisfy the inclusion/exclusion criteria of the trial and were administered the same control treatment as the internal controls. These individuals are viewed as a simple random sample from another hypothetical population, which we title the \textit{external control population}. No assumptions are made about the relationship between the external control and study populations. Collectively, the observed data are a simple random sample from a mixture population of the the study population and the external control population.

With these populations defined, it is now possible to describe the structure of the observed data. Let the outcome of interest be denoted by $Y$, the baseline covariates shared by the external control and sample populations by $Z$, and the binary treatment by $A$ ($a=1$ is the experimental treatment and $a=0$ is the control treatment, possibly a placebo, active control, or other intervention). For ease of exposition, we will refer to these dichotomous treatments as treatment and control, respectively. $D$ is an indicator for the data source ($d = 1$ for RCT and $d =0$ for external control). The observed data consists of $n$ independent and identically distributed samples of $O = (Y, Z, A, D)$ from the mixture population, where the $n_{rct}$ RCT and $n_{ec}$ external controls samples ($n_{rct} + n_{ec} = n$) have data structured as $(Y, Z, A, D=1)$ and $(Y, Z, A=0, D=0)$, respectively. For $a \in \{0, 1\}$, we posit the existence of potential outcomes $Y^a$ that represent the outcome that would have been observed under the the intervention to receive treatment $A = a$ \citep{rubin_causal_2005}. We make three assumptions about the RCT-design. \textbf{(1)}  \textit{Consistency of potential outcomes}: for each individual with $D = 1$ and for $a \in \{0,1\}$, $Y^a = Y$ if $A = a$. \textbf{(2)} \textit{Exchangeability of potential outcomes}: for $a \in \{0, 1\}$, $Y^a \perp A | D = 1$. \textbf{(3)} \textit{Positivity of treatment assignment}: $0 < \mathrm{Pr}(A = 1 | D = 1) < 1$.  Furthermore, it is assumed that there is no interference between the potential outcomes of different units. All three of these assumptions are expected to hold for well-defined treatments in a RCT. 

\subsection{Target Parameter}

A number of different parameters might be of scientific interest, including the ATE in the target population or a CATE. In this paper, we focus on a more easily interpreted causal target parameter: the ATE in the study population. Because this is frequently the treatment effect estimate of interest in a clinical trial, this is consistent with the target trial framework previously outlined. This treatment effect, $\tau$, is defined as the expected difference between the outcome if the study population were assigned the treatment versus if the study population were assigned the control: $\tau \equiv E[Y^1 - Y^0 | D = 1]$. When treatment effects are heterogeneous and the external control population differs from the study population, $\tau$ may differ from $E[Y^1 - Y^0 | D = 0]$ or $E[Y^1 - Y^0]$. Identification of these two parameters also requires assumptions beyond those for $\tau$. Because $D = 0$ implies $A = 0$ and internal study participants are randomized, we can equivalently define $\tau$ as $E[Y^1 - Y^0 | A = 1]$, which is the average treatment effect among the treated (ATT). 
\vspace{-0.5cm}

\section{Causal Identification}
\label{s:ident}

It is simple to verify that, under Assumptions 1-3, $\tau = E[Y | A = 1, D = 1] - E[Y | A = 0, D = 1]$, giving rise to the simple nonparametric estimator of the difference of mean outcomes between the treatment and internal control arms. Such identification is possible because, within the RCT ($D = 1$), patients are randomized to treatment. However, this identification result does not incorporate external controls, and thus no efficiency has been gained. The incorporation of external controls comes at the expense of randomization. Indeed, while $\mathrm{Pr}(A = 1 | D = 1)$ is under investigator control, $\mathrm{Pr}(A = 1) = \mathrm{Pr}(A = 1 | D = 1) \mathrm{Pr}(D = 1)$, and the assignment to $D$ is not randomized. To express $\tau$ in terms of external control data, we make assumptions about the similarity of $Y^0$ between the study and external control populations that aim to recover randomization.

\noindent \textbf{Assumption 4}. \textit{Mean-exchangeability of potential outcomes across populations}: There exists some $X \subset Z$ such that $E[Y^0 | D = 1, X=x] = E[Y^0 | D = 0, X=x]$ for all $x$ such that $p(x|D = 1) > 0$.

\noindent \textbf{Assumption 5}. \textit{Mean-consistency of potential outcomes among external controls}: $E[Y^0 | D = 0, X, A = 0] = E[Y | D = 0, X, A = 0]$

\indent Assumption 4 states that while $Y^0$ might systematically differ between the study and external control population due to differences in their distribution of $X$, once these differences are accounted for, the expectation of $Y^0$ is the same across populations. Investigators do not randomize individuals to their population $D$ (and thus their treatment $A$), but once $X$ is conditioned upon, the mean of $Y^0$ is independent of $D$, mimicking randomization. Assumption 5 is related but weaker than Assumption 2 in that it allows for inconsistencies in treatment delivery between internal and external controls as long as these inconsistencies are not systematic.

Equipped with these causal assumptions, the target parameter $\tau$ can be identified with the observed data. By conditioning, $\tau$ can be written as $E[Y^1 | D = 1] - E_{X | D = 1}[E[Y^0 | X, D = 1]]$. The first term, as noted earlier, is equivalent to $E[Y | A = 1]$ under Assumptions 1-3. The second term can be shown, using Assumptions 1-5, to be equal to $E_{X | D = 1}[E[Y | A = 0, X]]$.

Therefore, we have that
\begin{equation} \label{eq:ident_outcome}
    \tau = E[Y | A = 1] - E_{X | D = 1}[E[Y | A = 0, X]],
\end{equation} 
where $E[Y | A = 0, X]$ is over the mixture distribution of internal and external controls. 

Alternatively, $E[Y^0 | D = 1]$ can be expressed in terms of the probability of belonging to the study population. Letting $\pi_d(x) \equiv $ Pr$(D = 1 | X = x)$ and $\pi_a(x) \equiv $ Pr$(A = 1 | X = x, D = 1)$, we have that
\begin{equation} \label{eq:ident_prop}
\begin{split}
        &E_{X | D = 1}[E[Y^0 | X, D = 1]] =  \frac{1}{\mathrm{Pr}(D = 1)}\\
    &\times E \left[ \frac{(1-A)}{(1 - \pi_a(X)) \pi_d(X) + (1 - \pi_d(X))} \pi_d(X) Y  \right].
\end{split}
\end{equation}

The proposed identifiability criteria make no assumptions about the functional relationship of the expected potential outcomes under experimental treatment between the two data sources. Intuitively, no information from the external controls is informative about the outcome under the experimental treatment since no external controls receive the treatment. 
\vspace{-1cm}
\subsection{Graphical Criteria}

Assumptions 1-5 provide sufficient conditions under which the estimand $E[Y | A = 1] - E_{X | D = 1}[E[Y | A = 0, X] ]$ is (causally) identified. However, the assumptions are not constructive; a sufficient conditioning set of variables $X$ can only be inferred from the underlying causal model. Causal directed acyclic graphs (DAGS) present a powerful tool for allowing researchers to transparently communicate their qualitative knowledge and assumptions about the causal relationships among variables \citep{judea_pearl_causal_2009}. More recently, Single World Intervention Graphs (SWIGs) have unified causal DAGS and independence statements about potential outcomes \citep{richardson_single_2013},  such as Assumption 4. One of our objectives is to use causal graphical tools to characterize Assumption 4.

Previous work has outlined graphical conditions necessary to identify average treatment effects in the cases of generalizability \citep{dahabreh_generalizing_2019} and transportability \citep{bareinboim_general_2013}. Constructing causal graphs for this external control setting poses unique challenges. First, whereas a causal graph is typically defined with respect to a specific population, in the external control setting data is sampled from two, possibly distinct, populations. Furthermore, while the typical goal of causal graphs is to verify if $\{Y^0, Y^1\} \indep A | X$, in the external control setting the objective is instead to verify if $Y^0 \indep D | X$ (Assumption 4).

Motivated by these unique characteristics, we propose a variant of SWIGs based upon selection diagrams, which are modified causal graphs that determine the transportability of causal relations \citep{bareinboim_general_2013}. Let $\mathbf{G}$ denote a causal graph shared by the study and external control populations. A selection diagram $\mathbf{D}$ is created as follows:

\begin{itemize}
    \item Every edge in $\mathbf{G}$ is also an edge in $\mathbf{D}$;
    \item $\mathbf{D}$ contains an extra edge, $S_{Z_i} \longrightarrow Z_i$, whenever $p(z_i | v, D = 1) \neq p(z_i | v, D = 0)$ for some collection of variables $V \subset Z \backslash Z_i$;
    \item $\mathbf{D}$ contains an extra edge, $S_{Y} \longrightarrow Y$, whenever $p(y | z, D = 1) \neq p(y | z, D = 0)$.
\end{itemize}

These additional $S$ variables pinpoint differences across populations in the distribution of the $Z$ variables or in the distribution of $Y^0 | Z$. Importantly, $S$ variables identify differences in joint distributions, not just marginal ones. All variables with a causal effect on $Y^0$ should be included in the graph $\mathbf{G}$. See the Supporting Information for an extended discussion and the subsequent role of effect modification. From this selection diagram, we can create a SWIG by performing a node-split on $A$. Then, Assumption 4 is satisfied if there exists some set of variables $X \subset Z$ such that $Y^0 \indep S | X$ using the rules of d-separation. While this result is informative about which variables $X$ satisfy Assumption 4, it does not aid efficiency if $X$ completely separates the two populations such as when $X$ contains a binary feature that only takes on one of its values in the study population and its other value in the external control population.

\textbf{Theorem 1} \textit{Let the selection SWIG } $\mathbf{D}$ \textit{ be constructed as above. Then the following hold:}

\begin{enumerate}
    \item \textit{If } $Y^0 \perp S | X$, \textit{ where } $S$ \textit{ is the set of selection variables, then Assumption 4 holds, i.e. } $E[Y^0 | X, D = 1] = E[Y^0 | X, D = 1]$
    \item \textit{If } $Y^0 \not\perp S | X$, \textit{ then there exists some two distributions compatible with } $D$ \textit{ such that Assumption 4 does not hold, even if } $p(y^0 | z, D = 1) = p(y^0 | z, D = 0)$.
\end{enumerate}

\begin{figure*}
    \centering
\includegraphics[width = 7.0in]{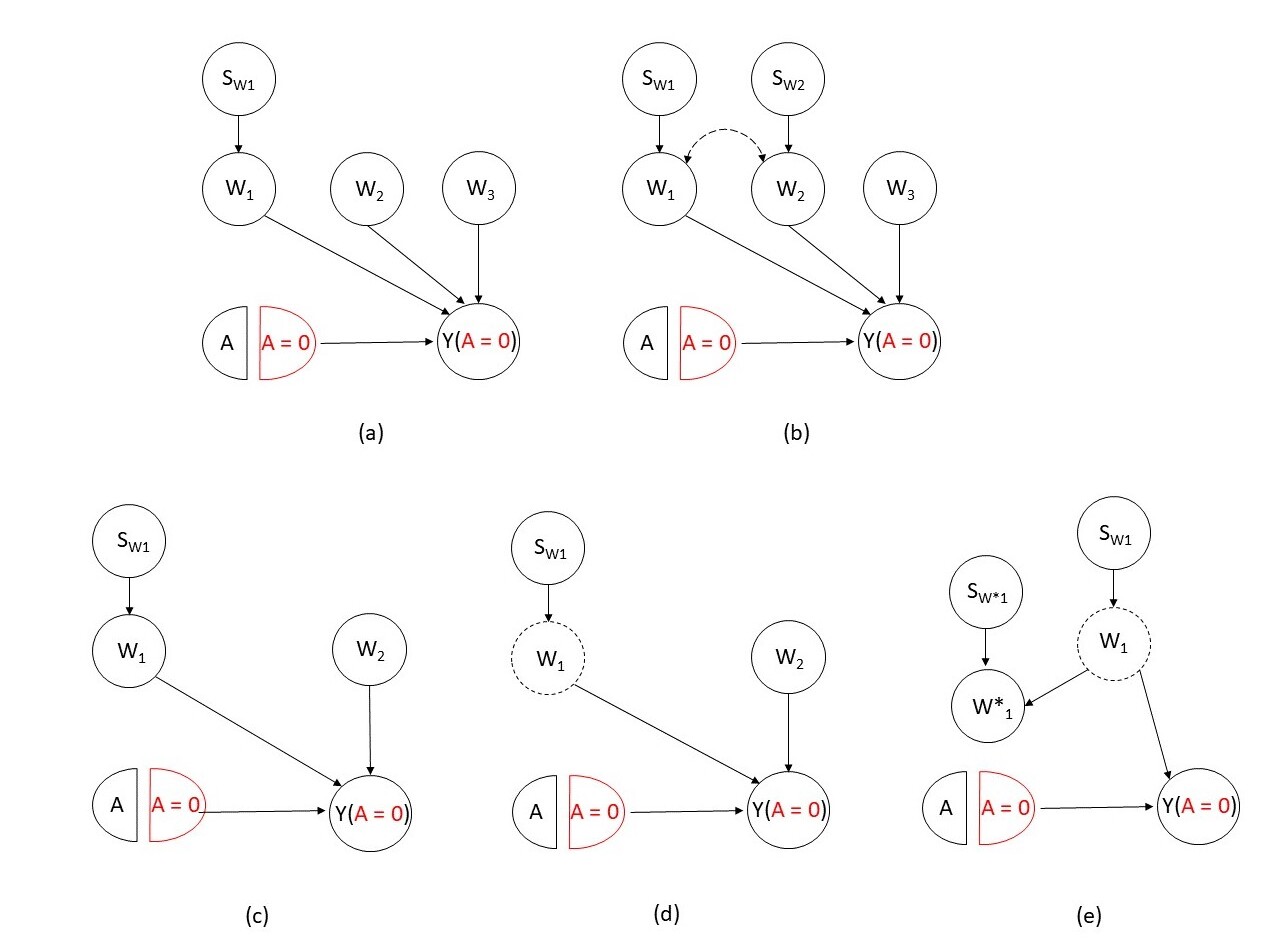}
\caption{Examples of Selection SWIGs. (a) By Theorem 1, $Y^0 \perp D | W_1$. (b) By Theorem 1, $Y^0 \perp D | W_1, W_2$. (c)  $W_1$ and $W_2$ are causes of $Y^0$, but only $W_1$ is imbalanced. Once $W_1$ is conditioned on, $S$ is independent of $Y^0$, and therefore $W_1$ is sufficient for adjustment. (d) When $W_1$ is unmeasured (dotted lines), causal bias is incurred. (e) $W_1$ should be adjusted for, but $W^*_1$ is the mis-measured variable available to investigators. The node $S_{W^*_1}$ pointing into $W_1^*$ represents the belief that the quality of $W^*_1$ (how well it measures $W_1$) differs between external controls and the study sample.}
\label{fig:bias}
\end{figure*}

As a concrete example, consider the setting in which it is hypothesized that the variables $\{Z_1, Z_2, Z_3 \}$ are causes of $Y^0$. Furthermore, $\mathrm{Pr}(Z_1 | D = 1) \neq \mathrm{Pr}(Z_1 | D = 0)$, $\mathrm{Pr}(z_2 | D = 1) = p(z_2 | D = 0)$, $\mathrm{Pr}(z_3 | D = 1) = \mathrm{Pr}(z_3 | D = 0)$, and $\{ Z_1, Z_2, Z_3 \}$ are mutually independent in both populations. It follows that $Y^0 \perp D | Z_1$. If the same marginal distributions hold but  $Z_1$ and $Z_2$ are not independent and $\mathrm{Pr}(Z_1, Z_2 | D = 1) \neq \mathrm{Pr}(Z_1, Z_2 | D = 0)$, we have that $Y^0 \not\perp D | Z_1$ but $Y^0 \perp D | Z_1, Z_2$. The selection SWIGs are depicted in Figure \ref{fig:bias}.

When the causal assumptions are violated, the causal parameter $\tau$ cannot be computed even with infinite data, a bias sometimes referred to as systematic bias \citep{hernan_causal_2023}. While sources of systematic bias in this setting, as well as heuristics to guard against it, have been previously proposed, the lack of a causal centering has led to criteria that can be vague or not directly target the underlying reason for the bias \citep{pocock_combination_1976}. We encourage the viewpoint that discussions of bias should focus on the extent to which the outlined causal assumptions are violated. Toward this goal, we highlight how common concerns with using external controls can threaten sound causal inference.

Geographic or temporal discrepancies between external controls patients and the study population can result in a cause of the outcome being imbalanced (hereafter meaning differing in distribution between the study population and the external control population). In the selection SWIGs, this is demarcated by the inclusion of a selection variable. It follows that the imbalance only threatens causal inference when the variable cannot be adjusted for, such as when it is unmeasured. Another concern, especially when the external control sample comes from a less quality-controlled source such as an EHR system, is measurement error. Because the external controls are all assigned the control, any form of mis-measurement will be differential with respect to the treatment.

\section{Estimation}
\label{s:estimation}
In what follows, let $\hat{m}_a(x)$ denote an estimator of $ m_a(x) \equiv E[Y | X = x, A = a]$, let $\hat{\pi}_a(x)$ be an estimator of the \textit{treatment propensity score} $\pi_a(x) \equiv \mathrm{Pr}(A = 1 | X = x, D = 1)$, and let $\hat{\pi}_d(x)$ be an estimator of the \textit{study propensity score} $\pi_d(x) \equiv \mathrm{Pr}(D = 1 | X = x)$.

Under the proposed causal assumptions, the target parameter $\tau$ is identified as the difference between two functionals:  $\mu_1 = E[Y | A = 1]$ and $\mu_0 = E_{X | D = 1}[E[Y | X, A = 0]]$. Because external controls are not administered the experimental treatment, they do not contribute to the estimation of this quantity. $\mu_0$ can be more efficiently estimated using both internal and external controls, through $\hat{m}_0(x)$, $\hat{\pi}(x)$, or both. Because these functions are not the target parameter but are required for its estimation, we refer to them as nuisance functions. When $X$ is high dimensional or contains continuous covariates, simple nonparametric estimates are no longer feasible, forcing the analyst to resort to modeling. Thus, differences in the distribution between internal and external controls, from a statistical perspective, heightens model dependency. In this section, we outline three classes of estimators: outcome model-based estimators that model $m_0(x)$, study-propensity score estimators that model $\pi_d(x)$, and efficient doubly-robust estimators that model both $m_0(x)$ and $\pi_d(x)$.

\subsection{Outcome Based Models}

The quantity $\mu_0$ averages the expected outcome of controls given baseline covariates $X$ over the distribution of $X$ in the study. Because it depends on the data only through $m_0(X)$ and the marginal distribution of $X$ in the study population, we can consider a simple plug-in estimator, where a regression model $\hat{m}_0(x)$ using the internal and external control data is fit and then averaged over the empirical distribution of $X$ from the study sample:
\begin{eqnarray*}
    \hat{\mu}_{0,om} =  \int \hat{m}_0(x) dP_{n, X|D=1} = (\sum_{i=1}^n D_i)^{-1} \sum_{i=1}^n D_i \hat{m}_0(X_i).
\end{eqnarray*}

In the causal inference literature, this estimator is frequently referred to as standardization or g-computation \citep{hernan_causal_2023, chatton_g-computation_2020}. The practical benefit of using external controls for this estimator is a reduction in variance of the estimate of $m_0(x)$. To estimate $m_0(x)$, one might assume a parametric model $m_0(x, \beta)$, where $\beta$ is a finite dimensional parameter, and use the estimator $m_0(x, \hat{\beta})$. However, this can introduce bias through over-smoothing, and therefore more flexible (nonparametric or machine learning) methods might be desirable. The consequences of using such estimators will be explored shortly.

\subsection{Study Propensity Score-Based Models (IPDW)}

As an alternative to modeling the outcome, $\pi_d(x)$ can be used to re-weight the outcomes of the controls so that their distribution of $X$ in this re-weighted population is the same as the treated. This idea is analogous to propensity score weighting methods in causal inference. The usual propensity score, $\pi(x) \equiv \mathrm{Pr}(A = 1 | x)$ is equal to $\pi_a(x) \pi_d(x)$. While $\pi_a(x)$ is known by trial-design, $\pi_d(x)$ is unknown, and so the treatment assignment mechanism across the entire sample is unknown, even though the hybrid trial incorporates data from an RCT.

The weighting method we consider is based upon the propensity formulation for $\mu_0$ (\ref{eq:ident_prop}). After estimating $\hat{\pi}_d(x)$ and $\hat{\pi}_a(x)$, we consider the plug-in estimator $\hat{\mu}_{0, ipdw} = n^{-1} \sum_{i=1}^n \hat{W}_i Y_i$, where
\begin{eqnarray}
    \hat{W}_i = \frac{n}{n_{rct}} \frac{1-A_i}{(1 - \hat{\pi}_a(X_i)) \hat{\pi}_d(X_i) + (1 - \hat{\pi}_d(X_i)} \hat{\pi}_d(X_i). \nonumber
\end{eqnarray}

While $\pi_a(x)$ is known through the trial design, it has been shown that estimating it (with the known model form) can lead to improved performance in finite-sample settings \citep{hahn_role_1998}. As before, one might assume a parametric model indexed by $\beta$ for $\pi_d(x)$. However, the conditional probability of belonging to the study population is likely a complicated function, and therefore more flexible models might be considered to limit the bias. Methods that handle extreme weights in $\hat{\mu}_{0, ipdw}$ or leverage $\hat{\pi}_d$ for matching instead of weighting are discussed in the Supporting Information.

\subsection{Efficient Estimators}

In the previous section, the proposed estimators incorporated a model for either $m_0(x)$ or $\pi_d(x)$. Heuristically, more precise estimators for $m_0(x)$ or $\pi_d(x)$ can produce a less variable estimator of $\tau$. However, this precision through incorporating external controls comes at the expense of additional causal assumptions (Assumptions 4-5) and an increased dependency on modeling. Since the goal of incorporating external controls is to increase efficiency in estimating $\tau$, a relevant aim is to understand how much variance reduction they can provide. 

Under results from the semiparametric efficiency theory literature, any regular and asymptotically linear (RAL) estimator for $\tau$ has an asymptotic variance of at least $\mathrm{B}_{\tau} \equiv E[\phi(O; \mathbb{P})^2]$, where $\phi(O; \mathbb{P})$ is the so-called \textit{efficient influence curve} for $\tau$ \citep{kosorok_introduction_2008}. This provides a bound on the efficiency of any RAL estimator. The efficient influence curve in this setting has been shown \citep{li_improving_2021} to be 

\begin{equation} 
\begin{split}
    \label{eq:influence_curve}
    \phi(O; \mathbb{P}) &= \frac{1}{q} \bigg[ D (m_1(X) - m_0(X) - \tau) + \frac{DA}{\pi_a(X)}(Y - m_1(X)) \\
    &- W(A, D, X) (Y - m_0(X)) \bigg],
\end{split}
\end{equation} where  $q \equiv \int \pi_d(x) d\mathbb{P}$, $r(X) \equiv \mathrm{Var}(Y^0 | X, D = 1)/\mathrm{Var}(Y^0 | X, D = 0)$ and where\begin{eqnarray}
    W(A, D, X) \equiv \frac{D(1-A) \pi_d(X) + (1-D) \pi_d(X) r(X)}{\pi_d(X)(1-\pi_a(X)) + (1-\pi_d(X))r(X)}. \nonumber
\end{eqnarray}

The authors showed that $\tilde{\mathrm{B}}_{\tau} -  \mathrm{B}_{\tau}$, where $\tilde{\mathrm{B}}_{\tau}$ is the efficiency bound in the model where no external controls are incorporated \citep{hahn_role_1998}, is strictly positive as long as there exists a nonzero measure set $A$ for which $\mathrm{Pr}(A | D = 1) \mathrm{Pr}(A | D = 0) > 0$. Furthermore, the difference is greatest when there is large overlap in the distribution of $X$ between the study and external control populations and when the noise in the distribution of $Y^0 | D = 0, X$ is small.  

We consider two types of estimators based upon the form of $\phi(O; P)$, where we write $\phi$ as a function of a distribution $P$ to highlight that it depends on the nuisance functions $m_1(x)$, $m_0(x)$, $\pi_d(x)$, and $r(x)$ as well as the target parameter $\tau$. If these functions were known, then the estimator $n^{-1} \sum_{i=1}^n \tilde{\phi}(O_i, \mathbb{P})$ would achieve the efficiency bound. Therefore, the first influence-curve based estimator, as proposed by \citep{li_improving_2021} and based upon an estimating equation, plugs in fitted models for the nuisance functions: \begin{equation*}
\begin{split}
    \hat{\tau}_{aipw} &= \frac{1}{n} \sum_{i=1}^n \frac{1}{\hat{q}}  \bigg[ D_i (\hat{m}_1(X_i) - \hat{m}_0(X_i)) \\
    &+ \frac{D_i A_i}{\hat{\pi}_a(X_i)}(Y_i - \hat{m}_1(X_i)) - \hat{W}(A_i, D_i, X_i) (Y_i - \hat{m}_0(X_i)) \bigg],
    \end{split}
\end{equation*}
where $\hat{q} = \frac{n_{rct}}{n}$ and\begin{eqnarray}
    \hat{W}(A, D, X) = \frac{D(1-A) \hat{\pi}_d(X) + (1-D) \hat{\pi}_d(X) \hat{r}(X)}{\hat{\pi}_d(X)(1-\hat{\pi}_a(X)) + (1-\hat{\pi}_d(X))\hat{r}(X)}. \nonumber
\end{eqnarray}

An alternative strategy is through targeted maximum likelihood estimation (TMLE) \citep{laan_targeted_2011}. While the two approaches here produce asymptotically equivalent estimators, TMLE is a plug-in estimator and thus respects the boundaries of the parameter space for all sample sizes, potentially leading to improved performance in small sample sizes. First, initial models $\hat{m}_1(x)$ and $\hat{m}_0(x)$ are fit to the data. Instead of using these models to form a simple plug-in estimator (in a fashion similar to $\hat{\mu}_{0, om}$), the models are fluctuated so that $n^{-1} \sum_{i=1}^n \phi(O_i, \hat{\mathbb{P}}^*) = 0$, implying that it, just like $\hat{\tau}_{aipw}$, solves the efficient influence curve estimating equation. Then, the TMLE estimator, $\hat{\tau}_{tmle}$ is just the plug-in estimator using the updated models $\hat{m}_1^*(x)$ and $\hat{m}_0^*(x)$: $\hat{\tau}_{tmle} = n^{-1} \sum_{i=1}^n \big( \hat{m}_1^*(X_i) - \hat{m}_0^*(X_i) \big)$.

\subsection{Inference}

When $\hat{m}_0$ is consistent for $m_0$, then $\hat{\mu}_{0,om}$ is consistent for $\mu_0$. Furthermore, when $\hat{\pi}_d(x)$ is consistent for $\pi_d(x)$, then $\hat{\mu}_{0, ipdw}$ is consistent for $\mu_0$. While this would seemingly encourage using flexible, data-adaptive methods to avoid misspecifying $m_0(x)$ and $\pi_d(x)$, doing so generally leads to slow rates of convergence and a poorly understood limiting distribution for $\hat{\tau}$. Under correctly specified parametric models (or more generally, models that belongs to a Donsker class \citep{kosorok_introduction_2008}), the nonparametric bootstrap can be used to construct confidence intervals for $\tau$ for either the outcome-modeling or IPDW approaches. 

\textbf{Proposition 2} \textit{Let $\hat{m}_0(x)$ and $\hat{\pi}_d(x)$ belong to a Donsker class $\mathcal{F}$ and assume that $\int (\hat{m}_0(x) - m_0(x))^2 dP_x = o_p(1)$ and $\int (\hat{\pi}_d(x) - \pi_d(x))^2 dP_x = o_p(1)$. Then the nonparametric bootstrap using outcome modeling or the IPDW approach is valid for $\mu_0$}.

The benefit of the outcome modeling and IPDW approaches is their relative simplicity and agnosticism about how $E[Y^1 | D - 1]$ is modeled. However, these methods are not efficient and face a trade-off between reducing bias through using flexible models and maintaining valid $\sqrt{n}$ inference. This motivates the usage of $\hat{\tau}_{aipw}$ and $\hat{\tau}_{tmle}$, which possess several desirable statistical properties. First, the models are doubly-robust in the sense that $\hat{\tau}$ converges in probability to $\tau$ if either $\hat{m}_0(x)$  converges in probability to $m_0(x)$ and $\hat{m}_1(x)$ converges in probability to $m_1(x)$ or if $\hat{\pi}_d(x)$ converges in probability to $\pi_d(x)$ \citep{li_improving_2021}. While this double-robustness is an attractive property, neither nuisance function is known in advance, leading to the plausible scenario in which parametric assumptions are too strong for both models. A more attractive property that we demonstrate is that, when cross-fitting is used, $\hat{\tau}_{aipw}$ and $\hat{\tau}_{tmle}$ are both efficient and asymptotically normally distributed \textbf{even} when the nuisance functions are estimated at slower rates, allowing for flexible estimation of these functions using popular machine learning methods that can limit bias. 

\textbf{Theorem 2} \textit{Suppose that the regularity conditions described in the Appendix hold. Then for} $\hat{\tau}_{dr}$, \textit{meaning either} $\hat{\tau}_{aipw}$ or $\hat{\tau}_{tmle}$, \textit{we have that} $\sqrt{n} (\hat{\tau}_{dr} - \tau) \longrightarrow N(0, E[\phi(O, \mathbb{P})])$ \textit{Therefore, $\hat{\tau}_{dr}$ is root-n consistent, semiparametric efficient, and asymptotically normal with asymptotically-valid confidence intervals given by} $\hat{\tau}_{dr} \pm 1.96 \sqrt{\hat{\mathrm{var}}(\phi(O, \hat{\mathbb{P}})) / n}$
\section{Analysis Pipeline}
\label{s:diag}

We envision an analysis pipeline for hybrid trials utilizing external controls based upon this causal inference framework. First, the target trial is defined, with precise definitions for the study population, inclusion criteria, and intervention of interest. Then, investigators encode their assumptions about the causal model and distributional differences between the two populations through a selection SWIG. Using the rules of d-separation on the selection SWIG, a set of variables $X$ can be determined that are sufficient to satisfy assumptions 4-5. An efficient estimator of the ATE can be constructed using either $\hat{\tau}_{aipw}$ or $\hat{\tau}_{tmle}$, with confidence intervals constructed as per Theorem 2.

While the selection SWIG provides a theoretical justification for Assumptions 4-5 based upon background knowledge of the causal process, empirical evidence can help weigh the validity of the assumptions. Assumption 4, when combined with the consistency assumptions, implies that $E[Y | A = 0, D = 1, X] = E[Y | A = 0, D = 0, X]$, a so-called "testable implication" of the causal assumptions. This is an assumption that two (nonparametric) regression functions, $m_0(x)$ and $m_1(x)$, are equal to each other (or, equivalently, that the regression function does not depend upon $D$). There is an expansive literature on testing this hypothesis, although many focus on the settings in which the dimension of $X$ is one, $X \in [0,1]$, or design points for $X$ are chosen in advance. We refer readers to \citep{racine_testing_2006}, \citep{lavergne_significance_2015}, or \citep{luedtke_omnibus_2019} for examples of applicable tests.

Alternatively, the study propensity score can be used to diagnose violations of the causal assumptions and to assess the degree to which the external controls differ from the study sample (and thus how dependent the results are upon modeling). When Assumption 4 holds, we expect that $E[Y | D, \pi_d(X), A = 0] = E[Y | \pi_d(X), A = 0]$. A simple diagnostic is therefore to compare the mean outcomes between internal and external controls with similar estimated propensity scores. Dissimilar mean outcomes between the two groups would suggest a violation of the causal assumptions. 

Greater distributional difference in $X$ between the internal and external controls results in heightened reliance upon the models for $m_0(x)$ and $\pi_d(x)$. A comparison of study propensity scores can be used to determine if the external controls are too different from the internal controls to reliably increase efficiency in estimation. In a parallel to methods used for generalizability, we define the propensity score difference as $\Delta_{\pi_d} = n_{rct}^{-1} \sum_{i=1}^n D_i \hat{\pi}_d(X_i) - n_{ec}^{-1} \sum_{i=1}^n (1-D_i) \hat{\pi}_d(X_i)$. When the external control population is the same as the study population, we expect this difference to be 0. Larger values indicate greater differences between the two populations. Related research in the context of using propensity scores in observational studies suggested that $\Delta_{\pi_d} > 0.25$ is suggestive of an over-dependence on modeling \citep{stuart_matching_2010}.

\section{Simulation Study}
\label{s:simulation}

\begin{table*}
    \centering
    \caption{Results (bias, mean-squared error, and 95\% confidence interval coverage) of the simulation. In Setting 1, all nuisance functions are estimated with correctly specified linear models. In Setting 2, an incorrectly specified linear model is used for $\pi_d(x)$ while Random Forests are used for $m_a(x)$. In Setting 3, incorrectly specified linear models are used for $m_a(x)$ while a Random Forest is used for $\pi_d(x)$. In Setting 4, incorrectly specified linear models are used for both $m_a(x)$ and $\pi_d(x)$. Both $m_a(x)$ and $\pi_d(x)$ are modeled with Random Forests in Setting 5}
\label{tab:sim_settings}
\resizebox{\textwidth}{!}{%
\begin{tabular}{lrrrrrrrrrrrrrrr}\toprule
& \multicolumn{3}{c}{$\hat{\tau}_{rct}$} & \multicolumn{3}{c}{$\hat{\tau}_{om}$} & \multicolumn{3}{c}{$\hat{\tau}_{ipdw}$} & \multicolumn{3}{c}{$\hat{\tau}_{aipw}$} & \multicolumn{3}{c}{$\hat{\tau}_{tmle}$}
\\\cmidrule(lr){2-4}\cmidrule(lr){5-7}\cmidrule(lr){8-10}\cmidrule(lr){11-13}\cmidrule(lr){14-16}
& Bias  & MSE & Cov.    & Bias  & MSE & Cov. & Bias  & MSE & Cov. & Bias  & MSE & Cov. & Bias  & MSE & Cov.\\\midrule
Setting 1	&	-1.8e-03	&	0.31	&	0.96	&	2.5e-04	&	\textbf{0.22}	&	0.93	&	-2.9e-03	&	0.24	&	0.93	&	-4.2e-04	&	0.23	&	0.95	&	\textbf{6.6e-06}	&	\textbf{0.22}	&	0.96 \\
Setting 2	&	\textbf{3.4e-02}	&	0.31	&	0.95	&	\	&	---	&	---	&	4.0e-01	&	0.38	&	0.86	&	2.0e-01	&	\textbf{0.25}	&	0.94	&	2.2e-01	&	\textbf{0.25}	&	0.94 \\
Setting 3	&	\textbf{3.6e-02}	&	0.34	&	0.95	&	4.1e-01	&	0.38	&	0.85	&	---	&	---	&	---	&	1.1e-01	&	\textbf{0.24}	&	0.96	&	6.8e-02	&	\textbf{0.24}	&	0.96 \\
Setting 4	&	\textbf{2.5e-02}	&	\textbf{0.34}	&	0.96	&	4.0e-01	&	0.39	&	0.85	&	4.0e-01	&	0.39	&	0.86	&	4.2e-01	&	0.42	&	0.86	&	4.1e-01	&	0.40	&	0.88 \\
Setting 5	&	\textbf{-4.7e-03}	&	0.32	&	0.95	&	2.2e-01	&	0.24	&	0.93	&	4.8e-02	&	0.22	&	0.94	&	5.8e-02	&	\textbf{0.21}	&	0.96	&	6.3e-02	&	\textbf{0.21}	&	0.96 \\\bottomrule
\end{tabular}}
\end{table*}

We perform a simulation study examining sensitivity to model mis-specification, confidence interval coverage, and power. Consistent with our SMA application, we simulate a trial with a continuous endpoint, 150 RCT patients randomized 2:1 to treatment, and 50 external controls. Full details of the data generation and models specifications are available in the Supporting Information.

To estimate $\tau$, we compare an RCT-only covariate-adjusted AIPW estimator ($\hat{\tau}_{rct}$) with the four proposed estimators ($\hat{\tau}_{om} \equiv \hat{\mu}_1 - \hat{\mu}_{0, om}$, $\hat{\tau}_{ipdw} \equiv \hat{\mu}_1 - \hat{\mu}_{0, ipdw}$, $\hat{\tau}_{aipw}$, and $\hat{\tau}_{tmle}$). Both doubly-robust approaches correspond to their cross-fit variants. During the simulations, 95\% confidence intervals were constructed for each method. For $\hat{\tau}_{om}$ and $\hat{\tau}_{ipdw}$, confidence intervals were constructed via the non-parametric bootstrap. Confidence intervals were generated for $\hat{\tau}_{aipw}$ and $\hat{\tau}_{tmle}$ using the closed-form formula from Theorem 1. All simulations are run for 1000 replicates.

First, we explore the role of model misspecification on bias, mean squared error, and confidence interval coverage via different combinations of correctly and incorrectly specified models for the nuisance functions (Table \ref{tab:sim_settings}).  
With correctly specified parametric models (Setting 1), all external control approaches are unbiased and exhibit less variability. However, performance gains are not guaranteed under departures from correctly-specified parametric models. In Settings 2 and 3 where one nuisance function is incorrectly specified, the singly-robust approaches ($\hat{\tau}_{om}$ and $\hat{\tau}_{ipdw}$) are biased. Conversely, the doubly-robust approaches (with one nuisance function estimated flexibly and the other missecified) have small degrees of bias but lower MSEs than $\hat{\tau}_{rct}$.  Furthermore, while double robustness offers no protection against bias when both nuisance functions are incorrectly specified (Setting 4), the data-adaptive estimation of both functions (Setting 5) leads to estimates with minimal bias and variability, highlighting the inferential gains that are possible through flexibly modeling the nuisance functions.

\begin{figure*}
\includegraphics[width=7.0in]{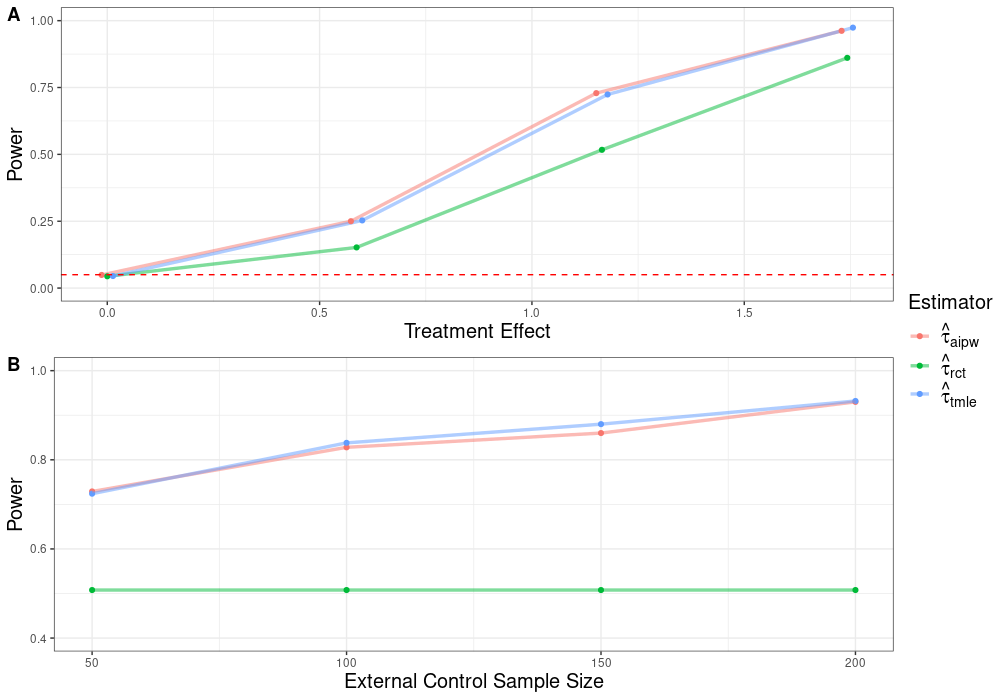}
\caption{(\textbf{A}) Power across different treatment effect sizes (points for $\hat{\tau}_{aipw}$ and $\hat{\tau}_{tmle}$ are jittered due to overlapping values). The dashed red line corresponds to the significance level of $0.05$ to demonstrate Type-1 error control under the null hypothesis. (\textbf{B}) Power across different external control sample sizes (with RCT sample sizes and treatment effect size held fixed). While the SMA application has an external control sample size around 50, other studies might have access to larger external control groups.}
\label{fig:power}
\end{figure*}

Next, to highlight applicability to clinical trials, we simulate power under varying degrees of treatment effect sizes and external control sample sizes (Figure \ref{fig:power}). We replicate Setting 5, modeling the (unknown) nuisance functions with Random Forests. Results are only depicted for $\hat{\tau}_{aipw}$ and $\hat{\tau}_{tmle}$ since valid confidence intervals are not generally attainable in this setting for $\hat{\tau}_{om}$ and $\hat{\tau}_{ipdw}$. When $\tau = 0$, both external control methods achieve satisfactory type-1 error control ($\leq \alpha = 5\%$). Across treatment effect sizes, power gains compared to $\hat{\tau}_{rct}$ ranged from 13\% to 65\%. Furthermore, while our SMA application only has a pool of $50$ external controls, more power gains are possible with larger sample sizes.

\section{SMA Example}
\label{s:SMA}

\begin{figure*}
\includegraphics[width=7.0in]{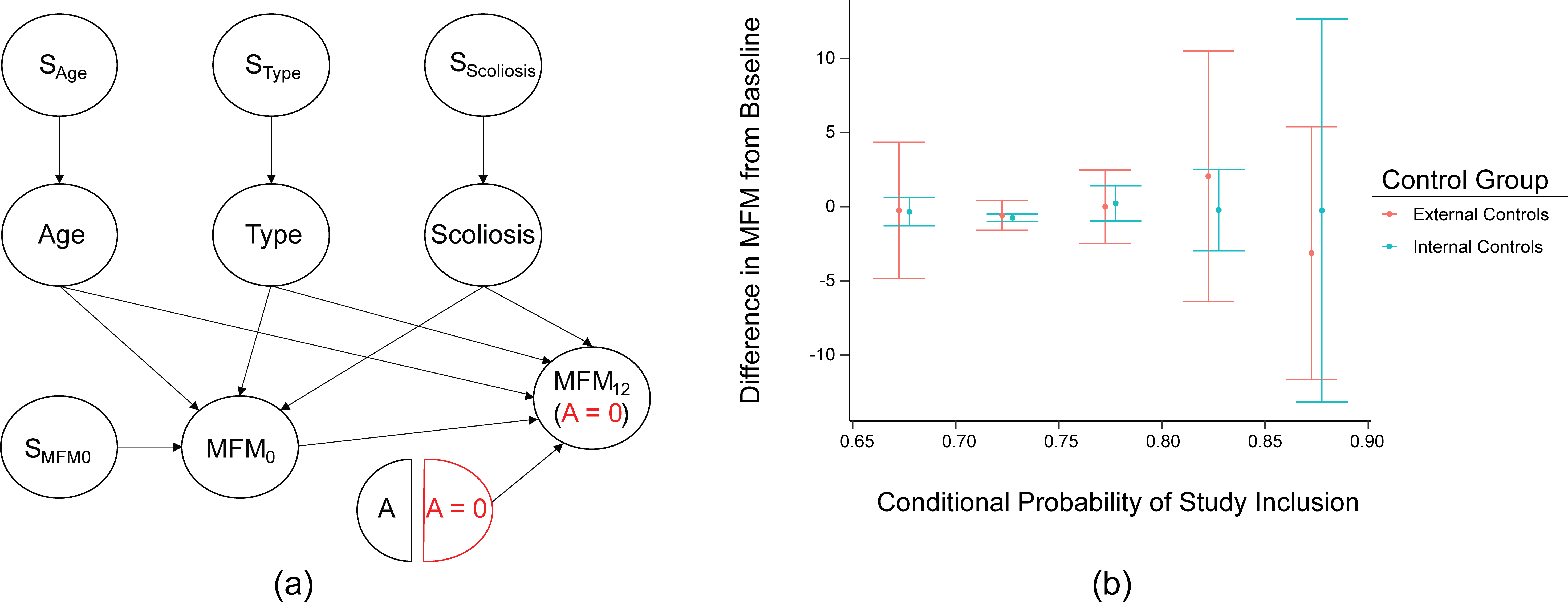}
\caption{\textbf{(a)} Selection Diagram for the SMA hybrid trial. MFM$_0$ and MFM$_{12}$ correspond to the MFM scores at baseline and 12 months, respectively. Type refers to the SMA Type and A is the treatment (risdiplam). \textbf{(b)} Diagnostic of Assumptions 4-5. Buckets were created based on $\hat{\pi}_d(x)$ in increments of 0.05. Error bars correspond to 95\% confidence intervals}
\label{fig:sma_example}
\end{figure*}

We demonstrate an implementation of the outlined causal inference framework for external controls in an application to a study of risdiplam for patients with spinal muscular atrophy (SMA). SUNFISH (NCT02908685) \citep{mercuri_safety_2022} was a two-part multi-site randomized placebo-controlled trial designed to investigate the efficacy of risdiplam.  In Part 2 of the study, 180 patients with Type 2 and non-ambulant Type 3 SMA were randomized 2:1 to receive risdiplam or control; the primary endpoint of interest was the change in Motor Function Measure (MFM) at 12 months ($\Delta_{MFM}$). Further details of the trial are provided in the Supporting Information. We explore a hybrid trial design that would incorporate external controls from the placebo arm of a Phase 2 trial of olesoxime (NCT01302600). After restricting to complete cases with Type 2 or non-ambulant Type 3 SMA between the ages of 2-25, the RCT and external control samples contain 159 and 48 observations, respectively. The target parameter of interest is defined as $\tau = E[\Delta_{MFM}^1 | D = 1] - E[\Delta_{MFM}^0 | D = 1]$. We are interested in testing $H_0: \tau = 0$ versus $H_1: \tau \neq 0$. We hypothesize the causal model depicted in Figure \ref{fig:sma_example} where no assumptions are made about differences in distribution of the covariates between the RCT and the external control populations. Based upon this selection diagram, $E[\Delta_{MFM}^0 | D, X] = E[\Delta_{MFM}^0 | X]$, where $X = ($Age, SMA Type, Scoliosis, MFM$_0)$.

 We estimate $\tau$ with just the RCT data ($\hat{\tau}_{rct}$) using AIPW and compare results to the four proposed estimators: $\hat{\tau}_{om}$, $\hat{\tau}_{ipdw}$, $\hat{\tau}_{aipw}$, and $\hat{\tau}_{tmle}$. For the doubly-robust methods, the nuisance functions are fit using Random Forests and cross-fitting while linear models are used for $\hat{\tau}_{om}$ and $\hat{\tau}_{ipdw}$. Confidence intervals are based upon the nonparametric bootstrap for $\hat{\tau}_{om}$ and $\hat{\tau}_{ipdw}$ and from the closed-form formula for $\hat{\tau}_{aipw}$ and $\hat{\tau}_{tmle}$.  
 All five methods provide similar estimates for $\tau$ and reject $H_0$. Overall, the confidence intervals for the four methods incorporating external controls have widths between 80\% and 93\% the width of the confidence interval using just the RCT data. 

\begin{table}
\centering
\caption{Estimates of the causal effect for the SMA application.}
\label{table:sma_results}
\begin{tabular}{@{}cccc@{}}
\hline
\multicolumn{1}{|l}{Method} & $\hat{\tau}$         & 95\% CI          &  \multicolumn{1}{l|}{p-value} \\ \hline
$\hat{\tau}_{rct}$ & 2.33 & (0.82, 3.85) & 0.002 \\
$\hat{\tau}_{om}$ & 1.93 & (0.71, 3.13) & $<$ 0.001 \\
$\hat{\tau}_{ipdw}$ & 1.79 & (0.54, 3.00) &  0.002 \\
$\hat{\tau}_{aipw}$ & 1.92 & (0.66, 3.18) & 0.003 \\
$\hat{\tau}_{tmle}$ & 1.80 & (0.54, 3.06) &  0.005 \\
& & & \\
\end{tabular}
\end{table}

To explore possible violations of Assumption 5, we plot (Figure \ref{fig:sma_example}) the means of $\Delta_{MFM}$ for controls grouped by $D$ and binning by $\hat{\pi}_d(x)$. Under Assumptions 5-6, we would expect that $E[\Delta_{MFM} | D = 1, A = 0, \pi_d(x)] = E[\Delta_{MFM} | D = 0, A = 0, \pi_d(x)]$. Overall, the results appear consistent with the assumptions. 

\section{Discussion}

While it is well-understood that deviations in characteristics and protocols between the RCT and external controls pose significant challenges to synthesizing information to estimate causal effects, the absence of a causal inference framework has hindered the clear communication of relevant concepts. The (causal) target parameter is rarely defined (\citep{li_improving_2021, zhou_incorporating_2021} are notable exceptions) and discussions of bias center on practical heuristics \citep{pocock_combination_1976, viele_use_2014}. Our proposed causal framework gives rise to an easy-to-interpret target parameter, encourages the embedding of assumptions and investigator knowledge into a causal graph, and allows bias to be explicitly defined as a property of the causal model or violation of the assumptions.

A limitation is the dependency upon Assumptions 4-5 to identify the causal effect, which can only be assessed after data collection and for which there might be little power to test empirically due to the small sample sizes commonplace in this setting. Furthermore, a null hypothesis of no difference might be of less interest than developing bounds based on the severity of the violations. While the methods discussed are applicable to continuous and binary outcomes, the extension to survival data would be an important contribution. 

\backmatter

\vspace{10cm}

\section*{Acknowledgements}

This work is partially supported by a grant from the FDA (NIH U01 FD007206, Kosorok, Pang, Valancius, Zhu) and a grant from the NIH (R01AI157758, Cole and Kosorok). We would like to thank our colleagues Tammy McIver and Wai Yin (Winnie) Yeung for their great comments and feedback.

\bibliographystyle{biom} \bibliography{references}

\label{lastpage}

\newpage
\newpage

\phantom{h}

\newpage



\section*{Supplementary material}

# Supporting Information for "A Causal Inference Framework for Leveraging External Controls in Hybrid Trials"

**Michael Valancius[1],[*], Herbert Pang[2], Jiawen Zhu[2],**

**Stephen R Cole[3], Michele Jonsson Funk[3], and Michael R Kosorok[1]**

[1]Department of Biostatistics, University of North Carolina at Chapel Hill, Chapel Hill, NC, U.S.A.

[2]Product Development Data Sciences, Genentech, South San Francisco, CA, U.S.A.

[3]Department of Epidemiology, University of North Carolina at Chapel Hill, Chapel Hill, NC, U.S.A.

[*]*email:* mval@email.unc.edu

## 1. Supporting Information

### 1.1 *Relationship to Transportability*

Unlike the causal inference task of transportability, the goal of a hybrid trial with external controls is not to transport inference from a study population to a target one. Instead, the task is somewhat reversed: external data are used for inference about the study population. Indeed, the target parameter previously discussed is the conventional target parameter in a traditional RCT. Despite these orthogonal targets of inference, the causal assumptions are similar in that they require similarity between the potential outcome distributions in the two populations.

Letting $S$ denote a binary indicator for selection into the study, the typical assumption made for transporting (or generalizing) results from a study to a target population is $E[Y^a|X, S = 1] = E[Y^a|X, S = 0]$ for $a \in \{0, 1\}$ (Dahabreh et al., 2019; Degtiar and Rose, 2023). Comparing this to Assumption 4 and equating $S$ with $D$, we observe that these assumptions are identical for $a = 0$, but no restrictions are placed on $Y^1$ since the external control cohort is never administered treatment $A = 1$.

### 1.2 *Alternative Identifiability Criteria*

The identification criteria listed, while sufficient for identifying $\tau$, are not exhaustive and alternative strategies might be explored depending on the context. The assumption that $E[Y^0|D = 1, X = x] = E[Y^0|D = 0, X = x]$ for all $x$ where $p(x|D = 1) > 0$ might be overly strong. More generally, it may be the case that $E[Y^0|D = 1, X] = E[Y|D = 0, X] + u(X)$, where $u(X)$ captures the effects of unmeasured prognostic factors. One simplifying assumption might be that $E[Y^0|D = 1, X] = E[Y^0|D = 1, X] + \mu$ for some constant $\mu$. This might be the case if there are baseline differences between internal and external controls not accounted for by $X$ but that these differences do not further impact $Y$ from study start to study conclusion.

1.2.1 *The Role of Effect Modification.* When constructing the causal graph **G**, all variables with a causal effect on the potential outcome $Y^0$ should be included. Implicit in this construction is that the set of variables depends not only on the outcome $Y$ but also on the choice of control treatment $A = 0$. A variable $X$ is titled an effect modifier on the additive scale if $E[Y^1|X = x] - E[Y^0|X = x]$ is not constant over $x$, implying an interaction between the treatment and $X$.

When transporting the causal effect from one population to another, all effect modifiers must be adjusted for (Dahabreh et al., 2019; Cole and Stuart, 2010). In a hybrid trial design with external controls, the form of the control $a = 0$ dictates which variables have an effect on $Y^0$. Investigators might believe that a placebo control necessitates less adjustment (in the form of $X$) than an active control.

To aid in clarity, consider the following hypothetical example. An investigator can choose among two control interventions: a placebo and an active control. The investigator is considering whether adjustments need to be made for two covariates that differ in distribution between the internal and external controls: age and sex. Sex is not believed to be have an effect on the outcome when patients are administered a placebo, but sex modifies the relationship between age and the outcome when the treatment is the active control. Figure 1 depicts this setting. If the placebo is used, only age is necessary for adjustment. If the active control is used, both age and sex are required. Ultimately, control interventions that are subject to greater degrees of effect modification pose heightened threats to the validity of the causal assumptions.

[Figure 1 about here.]

## 1.3 *Identification of $E[Y^0|D = 1, X]$*

Below, we show that, under Assumptions 1-5, $E[Y^0|D = 1, X] = E[Y|A = 0, X]$:

$$E[Y^0|D=1,X] = E[Y^0|D=1,X]\big(\Pr(D=1|A=0,X) + \Pr(D=0|A=0,X)\big)$$

$$= E[Y^0|D=1,X]\Pr(D=1|A=0,X)$$

$$+ E[Y^0|D=0,X]\Pr(D=0|A=0,X)$$

$$= E[Y^0|D=1,A=0,X]\Pr(D=1|A=0,X)$$

$$+ E[Y^0|D=0,A=0,X]\Pr(D=0|A=0,X)$$

$$= E[Y|D=1,A=0,X]\Pr(D=1|A=0,X)$$

$$+ E[Y|D=0,A=0,X]\Pr(D=0|A=0,X)$$

$$= E[Y|A=0,X]. \quad \square$$

The second equality holds by Assumption 4, the third equality holds by Assumption 2 (and because $D=0$ implies $A=0$), and the fourth equality holds by Assumptions 1 and 5.

### 1.4 *Identification of $E[Y^0|D=1,X]$ via the propensity formulation*

Below, we show that $E_{X|D=1}[E[Y^0|X,D=1]] = \frac{1}{\Pr(D=1)}E\left[\frac{(1-A)}{(1-\pi_a(X))\pi_d(X)+(1-\pi_d(X))}\pi_d(X)Y\right]$:

$$E_{X|D=1}[E[Y^0|X,D=1]] = \int E[Y|A=0,x]p(x|D=1)dx$$

$$= \frac{1}{\Pr(D=1)}\int E[Y|A=0,x]\pi_d(x)p(x)dx$$

$$= \frac{1}{\Pr(D=1)}\int\int y\frac{p(y|D=1,A=0,x)(1-\pi_a(x))\pi_d(x)}{(1-\pi_a(x))\pi_d(x)+(1-\pi_d(x))}\pi_d(x)p(x)dydx$$

$$+ \frac{1}{\Pr(D=1)}\int\int y\frac{p(y|D=0,A=0,x)(1-\pi_d(x))}{(1-\pi_a(x))\pi_d(x)+(1-\pi_d(x))}\pi_d(x)p(x)dydx$$

$$= \frac{1}{\Pr(D=1)}E\left[\frac{(1-A)D+(1-A)(1-D)}{(1-\pi_a(X))\pi_d(X)+(1-\pi_d(X))}\pi_d(X)Y\right]$$

$$= \frac{1}{\Pr(D=1)}E\left[\frac{(1-A)}{(1-\pi_a(X))\pi_d(X)+(1-\pi_d(X))}\pi_d(X)Y\right].$$

<div align="center">

1.5 *Proof of Theorem 1*

</div>

**Proof**

To see the first statement, we observe that there must be no $S_y$ present in the graph so that $p(y^0|W, D = 1) = p(y^0|W, D = 0)$. Let $X^c$ denote $W \backslash X$. Then we have the following, where in the third equality we make use of the observation from the previous line and the second step in the construction of $D$:

$$
\begin{aligned}
p(y^0|x, D = 1) &= \int p(y, x^c|x, D = 1)dx^c \\
&= \int p(y^0|x, x^c, D = 1)p(x^c|x, D = 1)dx^c \\
&= \int p(y^0|x, x^c, D = 0)p(x^c|x, D = 0)dx^c \\
&= p(y^0|x, D = 0) \qquad \square
\end{aligned}
$$

To see the second statement, let $S_{W_k}$ be a selection variable such that there is path from $S_{W_k}$ to $Y^0$ in $D$. Divide the support of $W_k$ into two distinct regions, $A_1$ and $A_0$ so that, with probability 1, $W_k$ takes on values in $A_1$ if $D = 1$ and values in $A_0$ if $D = 0$. Furthermore, let the distribution of $Y^0$ only depend on $W_k$ and divide the support of $Y$ into two distinct regions $B$ and $C$ such that $Y$ always has larger values in $B$ than in $C$. Finally, let the distribution of $Y^0$ be such that, with probability 1, $Y^0$ takes on values in $B$ if $W_k \in A_1$ and values in $C$ if $W_k \in A_0$. Clearly $E[Y^0|X, D = 1] > E[Y^0|X, D = 0]$ for all $X \subset W$. $\square$

<div align="center">

1.6 *Inverse-Propensity Weighting for Estimating the Average Treatment Effect on the Treated*

</div>

There is a direct connection between the proposed inverse propensity weighting method and the typical IPW method for estimating the ATT. The former is the empirical estimator of the following:

$$E[\frac{\Pr(D=1|X)}{\Pr(D=1)}\frac{(1-A)}{\Pr(A=0|X)}Y].$$

The latter, when estimating $E[Y^0|A=1$, is the empirical estimator of the following:

$$E[\frac{1}{\Pr(A=1)}\frac{\Pr(A=1|X)(1-A)}{\Pr(A=0|X)}Y].$$

We observe that the only difference is the ratios $\dfrac{\Pr(A=1|X)}{\Pr(A=1)}$ and $\dfrac{\Pr(D=1|X)}{\Pr(D=1)}$. However, these ratios are equivalent as long as $\Pr(A=1|X,D=1) = \Pr(A=1|D=1)$, as can be seen in the following:

$$
\begin{aligned}
\frac{\Pr(A=1|X)}{\Pr(A=1)} &= \frac{\Pr(A=1|X,D=1)\Pr(D=1|X) + \Pr(A=1|X,D=0)\Pr(D=0|X)}{\Pr(A=1|D=1)\Pr(D=1) + \Pr(A=1|D=0)\Pr(D=0)} \\
&= \frac{\Pr(A=1|X,D=1)\Pr(D=1|X) + 0}{\Pr(A=1|D=1)\Pr(D=1) + 0} \\
&= \frac{\Pr(D=1|X)}{\Pr(D=1)}.
\end{aligned}
$$

Therefore, we see that the inverse propensity score estimator estimates the same quantity as the inverse propensity method for calculating the ATT. The distinguishing characteristic, based on the factorization of $\Pr(A=1|X)$ into $\pi_a(X)\pi_d(X)$, is whether one estimates $\Pr(A=1|X)$ or $\pi_d(X)$.

### 1.7 *Handling Extreme Weights in* $\hat{\mu}_{0,ipdw}$

When the sample size is small, the weights used in $\hat{\mu}_{0,ipdw}$ can be extreme, leading to high variance and point estimates that might fall outside the support of $Y$. One remedy is to construct a bounded estimator that normalizes the weights:

$$\tilde{\mu}_{0,ipdw} = \left[\frac{1}{n}\sum_{i=1}^{n}\hat{W}_i\right]^{-1}\sum_{i=1}^{n}\hat{W}_iY_i.$$

### 1.8  *Matching Based on $\hat{\pi}_d$*

An alternative approach to balance the treated and control participants is matching. A common strategy when $X$ is high dimensional, as is likely in this setting, is to match treated and control units based upon the propensity score (Rosenbaum and Rubin, 1983). The propensity score $\pi(x)$ is a balancing score: $A \perp X | \pi(X)$ and therefore $Y^a \perp A | \pi(X)$ if $Y^a \perp A | X$ (Imbens and Rubin, 2015). A natural generalization to this setting is matching RCT individuals with external controls based upon the study propensity score $\pi_d(x)$. Analogous to the treatment propensity score, the study propensity score is a balancing score in the sense that $D \perp X | \pi_d(X)$, as shown in the following section. When combining $D \perp X | \pi_d(X)$ with Assumption 4, we have mean-exchangeability given $\pi_d(X)$. That is, $E[Y^0 | \pi_d(X), D] = E[Y^0 | \pi_d(X)]$ (details given in the next section).

Because $E[Y^0 | \pi_d(X)]$ is the same for internal and external controls for all $x$ with support in the study population, matching external controls to study participants with the same values of $\pi_d(X)$ and then taking the difference in outcomes between the treated and controls produces an unbiased estimate of the target parameter. However, in practice, $\pi_d(x)$ must be estimated and there will be no exact matches on $\hat{\pi}_d(X)$, leading to bias that depends on the accuracy of the estimator and the amount of discrepancy in treatment propensity scores of matched individuals (Abadie and Imbens, 2016).

### 1.9  *Balancing Score Property*

To show that $D \perp X | \pi_d(X)$, we will equivalently show that $\Pr(D = 1 | X, \pi_d(X)) = \Pr(D = 1 | \pi_d(X))$.

The left hand side of the equation is $\pi_d(X)$, since

$$\Pr(D = 1 | X, \pi_d(X)) = \Pr(D = 1 | X) = \pi_d(X).$$

Furthermore, by conditioning on $X$, we can see that the right hand side is also equal $\pi_d(X)$:

$$\Pr(D = 1|\pi_d(X)) = E[D|\pi_d(X)]$$

$$= E[E[D|\pi_d(X), X]|\pi_d(X)]$$

$$= E[\pi_d(X)|\pi_d(X)]$$

$$= \pi_d(X).$$

### 1.10  *Mean-exchangeability given $\pi_d(x)$*

To see this, we will now show that $E[Y^0|\pi_d(X), D] = E[Y^0|\pi_d(X)]$. This is shown in the following:

$$E[Y^0|\pi_d(X), D] = \int E[Y^0|x, \pi_d(x), d]p(x|\pi_d(x), d)dx$$

$$= \int E[Y^0|x, \pi_d(x)]p(x|\pi_d(x), d)dx$$

$$= \int E[Y^0|x, \pi_d(x)]p(x|\pi_d(x))dx$$

$$= E[Y^0|\pi_d(X)] \qquad \square$$

The second-to-last equality uses the fact that $p(x|\pi_d(x), d) = p(x|\pi_d(x))$ because $\Pr(D = 1|X = x, \pi_d(x)) = \Pr(D = 1|\pi_d(x))$.

### 1.11  *TMLE Implementation*

The target parameter $\tau$ is the contrast between $E[Y^1|D = 1]$ and $E[Y^0|D = 0]$. As before, we can identify $E[Y^1|D = 1]$ with $E_{X|D=1}[E[Y|A = 1, X]]$ and $E[Y^0|D = 1]$ with $E_{X|D=1}[E[Y|A = 0, X]]$. Therefore, a simple plug-in estimator of $\tau$ is given by:

$$\hat{\tau}_{plugin} = \frac{1}{n_1} \sum_{i=1}^{n} D_i\big(\hat{m}_1(X_i) - \hat{m}_0(X_i)\big)$$

A TMLE estimator of $\tau$ is also a plug-in estimator, but instead of using an original estimator $\hat{m}_a(x)$ of $E[Y|X = x, A = a]$, it updates $m_a(x)$ to produce $m_a^*(x)$, which is then

plugged into the above empirical average. Following Gruber and van der Laan (2010), we demonstrate how this update can be made when the outcome is continuous and bounded, although the results apply more generally to continuous outcomes if the bounds are taken to be that maximum and minimum observed values of $Y$.

Define $\tilde{Y} \equiv (Y - a)/(b - a)$, where $(a, b) = (\min(Y), \max(Y))$. Then, we can define $\tilde{\tau}$ as $E_{X|D=1}[E[\tilde{Y}|X, A = 1]] - E_{X|D=1}[E[\tilde{Y}|X, A = 0]]$. It follows that $\tilde{Y} \in [0, 1]$ and $\tau = \tilde{\tau}(b - a)$ so that inference for $\tilde{\tau}$ can be mapped to inference for $\tau$. Initial estimators $\hat{m}_0(x)$ and $\hat{m}_1(x)$ are constructed for $\tilde{m}_0(x) \equiv E[\tilde{Y}|X = x, A = 0]$ and $\tilde{m}_1(x) \equiv E[\tilde{Y}|X = x, A = 1]$. While these estimators traditionally have been constructed with SuperLearner (van der Laan et al., 2007), this need not be the case, and our simulations model these with Random Forests for a simpler baseline comparison of the methods. Furthermore, let $\hat{\pi}_d(x)$, $\hat{\pi}_a(x)$, and $\hat{r}(x)$ be estimators of $\pi_d(x)$, $\pi_a(x)$, and $r(x)$.

Define the clever covariate $h(D, A, X)$ as

$$h(D, A, X) \equiv \frac{1}{q}\left[\frac{DA}{\hat{\pi}_a(X)} - \frac{D(1-A)\hat{\pi}_d(X) + (1-D)(1-A)\hat{\pi}_d(X)\hat{r}(X)}{\hat{\pi}_d(X)(1 - \hat{\pi}_a(X)) + (1 - \hat{\pi}_d(X))\hat{r}(X)}\right].$$

We then fit a logistic regression model of $\tilde{Y}$ on $h(D, A, X)$ with offset given by the logit of $\hat{m}_A(X)$, ignoring that $\tilde{Y}$ is not binary. For each observation $O_i$, we can get predictions from this logistic regression model under the setting that $A = 1$ and the setting that $A = 0$. This defines our updated models $\hat{m}_1^*(x)$ and $\hat{m}_0^*(x)$. Then, our estimator for $\tau$ is given by:

$$\hat{\tau}_{tmle} = \left(\frac{1}{b - a}\right)\frac{1}{n_1}\sum_{i=1}^{n} D_i\big(\hat{m}_1^*(X_i) - \hat{m}_0^*(X_i)\big).$$

### 1.12  *Semiparametric Theory*

The ideas in the following are based upon a growing literature on the property of causal inference estimators that leverage machine learning methods to non-parametrically estimate

nuisance functions (Díaz, 2020; Kennedy, 2023; Hines et al., 2022; Chernozhukov et al., 2018).

More formally, we can think of $\tau$ as a map from $\mathcal{P}$ to $\mathbb{R}$, where $\mathcal{P}$ is a semiparametric model containing the true distribution $\mathbb{P}$ and whose only restriction is that $E[Y^0|X, D = 1] = E[Y^0|X, D = 0]$ almost surely.[1] As before, we can express $\tau(P)$ as the following:

$$\tau(P) = E_{P_{X|D=1}}[E_P[Y|A = 1, X]] - E_{P_{X|D=1}}[E_P[Y|A = 0, X]].$$

We note that this expression depends upon $P$ only through $E_P[Y|A = 0, X]$, $P_{X|D=1}$, and $E_P[Y|A = 1, X]$. In this semiparametric model, the efficient influence curve for the parameter $\tau(P)$ is given by Li et al. (2021) to be

$$\phi(O; \mathbb{P}) = \frac{1}{q}\left[D(m_1(X) - m_0(X) - \tau) + \frac{DA}{\pi_a(X)}(Y - m_1(X)) - W(A, D, X)(Y - m_0(X))\right], \quad (1)$$

where $q \equiv \int \pi_d(x)d\mathbb{P}$, $r(X) \equiv \mathrm{Var}(Y^0|X, D = 1)/\mathrm{Var}(Y^0|X, D = 1)$ and where

$$W(A, D, X) \equiv \frac{D(1 - A)\pi_d(X) + (1 - D)\pi_d(X)r(X)}{\pi_d(X)(1 - \pi_a(X)) + (1 - \pi_d(X))r(X)}.$$

To analyze the doubly-robust estimators when using machine learning models for the nuisance functions, we make use of the Von-Mises expansion (Vaart, 1998; Kennedy, 2023; Hines et al., 2022):

$$\tau(\overline{P}) - \tau(P) = \int \phi(o, \overline{P})d(\overline{P} - P)(o) + R_2(\overline{P}, P).$$

where $P\phi(O, P) = 0$, $P\phi(O, P)^2 < \infty$, and $R_2(\overline{P}, P)$ is a second order remainder term. We wish to consider the performance of an estimator that substitutes $\hat{\mathbb{P}}$ for $\mathbb{P}$, where $\hat{\mathbb{P}}$ denotes a model in which we have used estimators for $E_P[Y|A - 0, X]$, $P_{X|D=1}$, and $E_P[Y|A = 1, X]$.

---

[1] While $\pi_a(x)$ is also known, it is ancillary to the estimation of $\tau$ in the sense that the efficiency bound for $\tau$ is the same regardless of whether of not $\pi_a(x)$ is known.

We first consider the decomposition of the plug-in estimator that plugs in $\hat{\mathbb{P}}$ for $\mathbb{P}$. Based upon the von-Mises expansion, we have that it can be expressed in the following way:

$$\tau(\hat{\mathbb{P}}) - \tau(\mathbb{P}) = -\mathbb{P}\phi(O, \hat{\mathbb{P}}) + R_2(\hat{\mathbb{P}}, \mathbb{P})$$

The term $-\mathbb{P}\phi(O, \hat{\mathbb{P}})$ is a first-order bias term, and a common way of removing this bias is through adding an estimate of it: $\mathbb{P}_n\phi(O, \hat{\mathbb{P}})$. This new estimator, $\tau(\hat{\mathbb{P}}) + \mathbb{P}_n\phi(O, \hat{\mathbb{P}})$, is exactly equal to $\hat{\tau}_{aipw}$:

$$
\begin{aligned}
\tau(\hat{\mathbb{P}}) + \mathbb{P}_n\phi(O, \hat{\mathbb{P}}) &= \big(\sum_{i=1}^n D_i\big)^{-1} \sum_{i=1}^n D_i\big(\hat{m}_1(X_i) - \hat{m}_0(X_i)\big) + \sum_{i=1}^n \phi(O_i, \hat{\mathbb{P}}) \\
&= \big(\sum_{i=1}^n D_i\big)^{-1} \sum_{i=1}^n D_i\big(\hat{m}_1(X_i) - \hat{m}_0(X_i)\big) - \frac{1}{q} \sum_{i=1}^n D_i\big(\hat{m}_1(X_i) - \hat{m}_0(X_i)\big) \\
&\quad + \frac{1}{n}\frac{1}{\hat{q}} \sum_{i=1}^n \bigg[ D_i\big(\hat{m}_1(X_i) - \hat{m}_0(X_i)\big) + \frac{D_i A_i}{\hat{\pi}_a(X_i)}(Y_i - \hat{m}_1(X_i)) \\
&\quad - \hat{W}(A_i, D_i, X_i)(Y_i - \hat{m}_0(X_i)) \bigg] \\
&= \frac{1}{n}\frac{1}{\hat{q}} \sum_{i=1}^n \bigg[ D_i\big(\hat{m}_1(X_i) - \hat{m}_0(X_i)\big) + \frac{D_i A_i}{\hat{\pi}_a(X_i)}(Y_i - \hat{m}_1(X_i)) \\
&\quad - \hat{W}(A_i, D_i, X_i)(Y_i - \hat{m}_0(X_i)) \bigg] \\
&= \hat{\tau}_{aipw}
\end{aligned}
$$

To observe the asymptotic equivalence between $\hat{\tau}_{aipw}$ and $\hat{\tau}_{tmle}$, we note that $\hat{\tau}_{aipw}$ is defined by solving the estimating equation $\mathbb{P}_n\{\phi(O, \hat{\mathbb{P}}, \tau)\} = 0$. Conversely, $\hat{\tau}_{tmle}$ is defined as $\tau(\hat{\mathbb{P}}^*) = \tau(\hat{\mathbb{P}}^*) + \mathbb{P}_n\{\phi(O, \hat{\mathbb{P}}^*)\}$ since $\hat{\mathbb{P}}^*$ is defined so that $\mathbb{P}_n\{\phi(O, \hat{\mathbb{P}}^*)\} = 0$.

We consider the following decomposition based on the von-Mises expansion (Kennedy, 2023):

$$\hat{\tau}_{dr} - \tau(\mathbb{P}) = \tau(\hat{\mathbb{P}}) + \mathbb{P}_n \phi(O, \hat{\mathbb{P}}) - \tau(\mathbb{P}) \tag{2}$$

$$= (\mathbb{P}_n - \mathbb{P})\{\phi(O, \hat{\mathbb{P}})\} + R_2(\hat{\mathbb{P}}, \mathbb{P}) \tag{3}$$

$$= (\mathbb{P}_n - \mathbb{P})\{\phi(O, \mathbb{P})\} + (\mathbb{P}_n - \mathbb{P})\{\phi(O, \hat{\mathbb{P}}) - \phi(O, \mathbb{P})\} + R_2(\hat{\mathbb{P}}, \mathbb{P}). \tag{4}$$

In the above, $\hat{\tau}_{dr}$ stands for either $\hat{\tau}_{aipw}$ or $\hat{\tau}_{tmle}$, where it is understood that $\hat{\mathbb{P}}$ is $\hat{\mathbb{P}}^*$ when referring to $\hat{\tau}_{tmle}$.

By the central limit theorem, the first term converges to a normally distributed random variable with mean 0 and variance equal to the efficiency bound (scaled by $1\backslash n$). Hence, if the other terms are asymptotically negligible compared to the first term ($o_{\mathbb{P}}(1\backslash\sqrt{n})$), then $\hat{\tau}_{dr}$ is $\sqrt{n}$−consistent, asymptotically normal, and semiparametric efficient.

We first show how to control the second term (often termed the empirical process term). While Donsker conditions can be used to show this term is asymptotically negligible, we opt for cross-fitting (sometimes called sample-splitting) due to its simplicity and weaker conditions. In sample splitting, the full data, $O^N = (O_1, ..., O_K)$ is randomly split into $K$ (approximately) even sized folds. For each fold $k$ of size $N_k$, the nuisance functions of the influence curve ($m_0(x)$, $\pi_d(x)$, etc.) are estimated using only observations not contained in the $k$th fold, producing an estimator $\hat{\mathbb{P}}_{-k}$ of $\mathbb{P}$. Altogether, this produces $K$ estimators: $(\hat{\mathbb{P}}_{-1}, ..., \hat{\mathbb{P}}_{-K})$. Then $\hat{\tau}_{dr}$ is constructed by first fitting $K$ estimators $\hat{\tau}_{dr}^k$ where, for the $k$th estimator, the influence curve takes $\hat{\mathbb{P}}_{-k}$ instead of $\hat{\mathbb{P}}$ as its argument and is averaged over the data in the $k$th fold. Then, $\hat{\tau}_{dr}$ is the weighted average of these $K$ estimators $\hat{\tau}_{dr}^k$, with the weights corresponding to the proportion of the total observations in the fold. More details can be found in [Kennedy (2023)](#).

**Lemma 1** ([Kennedy (2023)](#)). Let $\hat{f}(o)$ be a function estimated from a sample $O^N = (O_{n+1}, ..., O_N)$, and let $\mathbb{P}_n$ denote the empirical measure over $(O_1, ..., O_n)$, which is independent of $O^N$. Then:

$$(\mathbb{P}_n - \mathbb{P})(\hat{f} - f) = O_{\mathbb{P}}\left(\frac{||\hat{f}_n - f||}{\sqrt{n}}\right)$$

Therefore, to show that $(\mathbb{P}_n - \mathbb{P})\{\phi(O, \hat{\mathbb{P}}) - \phi(O, \mathbb{P})\}$ is asymptotically negligible, it suffices to show that $\phi(O, \hat{\mathbb{P}}_{-k})$ converges to $\phi(O, \mathbb{P})$ in $L_2(\mathbb{P})$ norm for each $k$:

$$||\phi(\hat{\mathbb{P}}_{-k}) - \phi(\mathbb{P})|| \equiv \left(\int \big(\phi(o, \hat{\mathbb{P}}_{-k}) - \phi(o, \mathbb{P})\big)^2 d\mathbb{P}(o)\right)^{1/2} = o_{\mathbb{P}}(1).$$

Below, we present conditions for which this is true. For readability, we omit the distinction between $\hat{\mathbb{P}}_{-k}$ and $\hat{\mathbb{P}}$.

**Lemma 2** Suppose that:

- $||\hat{m}_0(x) - m_0(x)||$, $||\hat{m}_1(x) - m_1(x)||$, $||\hat{\pi}_a(x) - \pi_a(x)||$, $||\hat{\pi}_d(x) - \pi_d(x)||$, and $||\hat{r}(x) - r(x)||$ are all $o_{\mathbb{P}}(1)$.

- $\hat{\pi}_d(x)$ and $\pi_d(x)$ are greater than some $\epsilon_d > 0$ with probability 1

- $\hat{\pi}_a(x)$ and $\pi_a(x)$ are greater than some $\epsilon_a > 0$ and less than $1 - \epsilon_a < 1$ with probability 1

- $\beta(x) \equiv \pi_d(x)\big(1 - \pi_a(x)\big) + \big(1 - \pi_d(x)\big)r(x)$ and $\hat{\beta}(x) \equiv \hat{\pi}_d(x)\big(1 - \hat{\pi}_a(x)\big) + \big(1 - \hat{\pi}_d(x)\big)\hat{r}(x)$ are greater than some $\delta > 0$ with probability 1.

- $0 < r(X) < R$ with probability 1.

- $A(Y - m_1(X)) \leqslant M_1$ and $(1 - A)(Y - m_0(X)) \leqslant M_0$ with probability 1.

Then $||\phi(\hat{\mathbb{P}}) - \phi(\mathbb{P})||^2 = o_{\mathbb{P}}(1)$.

**Proof**

$$\begin{aligned}
\phi(\hat{\mathbb{P}}) - \phi(\mathbb{P}) &= \frac{1}{q}\Big[D\big(\hat{m}_1(x) - m_1(x)\big) - D\big(\hat{m}_0(x) - m_0(x)\big) + \frac{DA}{\hat{\pi}_a(x)}\big(Y - \hat{m}_1(x)\big) \\
&\quad - \frac{DA}{\pi_a(x)}\big(Y - m_1(x)\big) - \hat{W}(A, D, X)\big(Y - \hat{m}_0(x)\big) + W(A, D, X)\big(Y - m_0(x)\big)\Big] \\
&\equiv \frac{1}{q}\big[A - B\big].
\end{aligned}$$

Evaluating these terms separately, we obtain that:

$$\frac{1}{q}\big[A\big] = \frac{1}{q}D\big[\hat{m}_1(x) - m_1(x) - \frac{A}{\hat{\pi}_a(x)}\hat{m}_1(x) + \frac{A}{\pi_a(x)}m_1(x) + \frac{A}{\hat{\pi}_a(x)}Y - \frac{A}{\pi_a(x)}Y\big]$$

$$= \frac{1}{q}D\big[\big(\hat{m}_1(x) - m_1(x)\big)\big(1 - \frac{A}{\pi_a(x)}\big) + \frac{A}{\hat{\pi}_a(x)\pi_a(x)}\big(Y - \hat{m}_1(x)\big)\big(\pi_a(x) - \hat{\pi}_a(x)\big)\big].$$

Similarly, we find that:

$$\frac{1}{q}\big[B\big] = \frac{1}{q}\big[D\hat{m}_0(x) - Dm_0(x) + \hat{W}(A,D,X)\big(Y - \hat{m}_0(x)\big) - W(A,D,X)\big(Y - m_0(x)\big)\big]$$

$$= \frac{1}{q}\big[\big(\hat{m}_0(x) - m_0(x)\big)\big(D - W(A,D,X)\big) + \hat{W}(A,D,X)\big(Y - \hat{m}_0(x)\big)$$

$$- W(A,D,X)\big(Y - \hat{m}_0(x)\big)\big]$$

$$= \frac{1}{q}\big[\big(\hat{m}_0(x) - m_0(x)\big)\big(D - W(A,D,X)\big) + \big(Y - \hat{m}_0(x)\big)\big(\hat{W}(A,D,X) - W(A,D,X)\big)\big].$$

By the sub-additive property of norms, we then have that:

$$||\phi(\hat{\mathbb{P}}) - \phi(\mathbb{P})|| \leqslant \frac{1}{q}\big[||D\big(\hat{m}_1(x) - m_1(x)\big)\big(1 - \frac{A}{\pi_a(x)}\big)|| + ||\frac{A}{\hat{\pi}_a(x)\pi_a(x)}\big(Y - \hat{m}_1(x)\big)\big(\pi_a(x) - \hat{\pi}_a(x)\big)||$$

$$+ ||\big(\hat{m}_0(x) - m_0(x)\big)\big(D - W(A,D,X)\big)||$$

$$+ ||\big(Y - \hat{m}_0(x)\big)\big(\hat{W}(A,D,X) - W(A,D,X)\big)||\big]$$

$$\leqslant \frac{1}{q}\big[\frac{1}{\epsilon_a}||\hat{m}_1(x) - m_1(x)|| + ||\frac{M_1}{\epsilon_a^2}\big(\pi_a(x) - \hat{\pi}_a(x)\big)|| + \frac{R+1}{\epsilon_a}||\hat{m}_0(x) - m_0(x)||$$

$$+ ||M_0\big[D\big(\frac{\hat{\pi}_d(x)}{\hat{\beta}(x)} - \frac{\pi_d(x)}{\beta(x)}\big) + (1-D)\big(\frac{\hat{\pi}_d(x)\hat{r}(x)}{\hat{\beta}(x)} - \frac{\pi_d(x)r(x)}{\beta(x)}\big)\big]||\big]$$

$$= o_{\mathbb{P}}(1) + \frac{1}{q}||M_0\big[D\big(\frac{\hat{\pi}_d(x)}{\hat{\beta}(x)} - \frac{\pi_d(x)}{\beta(x)}\big) + (1-D)\big(\frac{\hat{\pi}_d(x)\hat{r}(x)}{\hat{\beta}(x)} - \frac{\pi_d(x)r(x)}{\beta(x)}\big)\big]||.$$

Therefore, it only remains to show that the second term, which comes from the simplification of $\hat{W}(A,D,X) - W(A,D,X)$, is $o_{\mathbb{P}}(1)$. We show this by evaluating the two parts of the term separately. The first part can be decomposed into terms that are all $o_{\mathbb{P}}(1)$ in the following way:

$$\frac{\hat{\pi}_d(x)}{\hat{\beta}(x)} - \frac{\pi_d(x)}{\beta(x)} = \frac{\hat{\pi}_d(x)\beta(x) - \pi_d(x)\hat{\beta}(x)}{\beta(x)\hat{\beta}(x)}$$

$$= \frac{\pi_d(x)\hat{\pi}_d(x)\big(\hat{\pi}_a(x) - \pi_a(x)\big) + \big(\hat{\pi}_d(x) - \pi_d(x)\big) + \pi_d(x)\hat{\pi}_d(x)\big(\hat{r}(x) - r(x)\big)}{\beta(x)\hat{\beta}(x)}.$$

The second part has a similar decomposition:

$$\frac{\hat{\pi}_d(x)\hat{r}(x)}{\hat{\beta}(x)} - \frac{\pi_d(x)r(x)}{\beta(x)} = \frac{\hat{\pi}_d(x)\hat{r}(x)\beta(x) - \pi_d(x)r(x)\hat{\beta}(x)}{\beta(x)\hat{\beta}(x)}$$

$$= \frac{\pi_d(x)\hat{\pi}_d(x)\big(\hat{r}(x) - r(x)\big) + \pi_d(x)\hat{\pi}_d(x)\big(\hat{\pi}_a(x)r(x) - \pi_a(x)\hat{r}(x)\big)}{\beta(x)\hat{\beta}(x)}$$

$$+ \frac{r(x)\hat{r}(x)\big(\hat{\pi}_d(x) - \pi_d(x)\big)}{\beta(x)\hat{\beta}(x)}.$$

Combining these results, we obtain that:

$$||M_0\big[D\big(\frac{\hat{\pi}_d(x)}{\hat{\beta}(x)} - \frac{\pi_d(x)}{\beta(x)}\big) + (1 - D)\big(\frac{\hat{\pi}_d(x)\hat{r}(x)}{\hat{\beta}(x)} - \frac{\pi_d(x)r(x)}{\beta(x)}\big)\big]||$$

$$\leqslant \frac{1}{\delta^2}\big[||\hat{\pi}_a(x) - \pi_a(x)|| + ||\hat{\pi}_d(x) - \pi_d(x)|| + ||\hat{r}(x) - r(x)||$$

$$||\hat{r}(x) - r(x)|| + ||\hat{\pi}_a(x)r(x) - \pi_a(x)\hat{r}(x)|| + R^2||\hat{\pi}_d(x) - \pi_d(x)||\big]$$

$$= o_{\mathbb{P}}(1) + \frac{1}{\delta^2}\big[||r(x)\big(\hat{\pi}_a(x) - \pi_a(x)\big)|| + ||\pi_a(x)\big(r(x) - \hat{r}(x)\big)||\big]$$

$$= o_{\mathbb{P}}(1) \quad \square$$

**Corollary 1**: Let the assumptions given by Lemma 2 hold for $\hat{\mathbb{P}}_{-k}$ for each $k$. Then the cross-fit doubly robust estimator $\hat{\tau}_{dr}$ has the property that:

$$\hat{\tau}_{dr} - \tau(\mathbb{P}) = (\mathbb{P}_n - \mathbb{P})\{\phi(O, \mathbb{P})\} + R_2^*(\hat{\mathbb{P}}, \mathbb{P}) + o_{\mathbb{P}}(1/\sqrt{n}) \qquad (5)$$

where $R_2^*(\hat{\mathbb{P}}, \mathbb{P}) = \sum_{k=1}^{K} \frac{N_k}{n} R_2(\hat{\mathbb{P}}_{-k}, \mathbb{P})$

We now turn our attention to the third term: $R_2^*(\hat{\mathbb{P}}, \mathbb{P})$

**Lemma 3**: Let $\hat{\mathbb{P}}$ be defined by estimators $\hat{\pi}_a(x)$, $\hat{\pi}_d(x)$, $\hat{m}_1(x)$, $\hat{m}_0(x)$, and $\hat{r}(x)$. Furthermore, let $\mathbb{P}D = \hat{\mathbb{P}}D = \dfrac{n_1}{n}$. Then we have the following:

$$
R_2(\hat{\mathbb{P}}, \mathbb{P}) = \int \Bigg( (m_0(x) - m_1(x)) + (\hat{m}_1(x) - \hat{m}_0(x)) + \frac{\pi_a(x)}{\hat{\pi}_a(x)}(m_1(x) - \hat{m}_1(x))
$$
$$
- \frac{\hat{\pi}_d(x)}{\pi_d(x)}(m_0(x) - \hat{m}_0(x)) \frac{\pi_d(x)(1 - \pi_a(x)) + (1 - \pi_d(x))\hat{r}(x)}{\hat{\pi}_d(x)(1 - \hat{\pi}_d(x)) + (1 - \hat{\pi}_d(x))\hat{r}(x)} \Bigg) f(x|d=1)dx
$$

**Proof**:

By the Von-Mises expansion and the form of $\phi(O, P)$, we have that:

$$
R_2(\hat{\mathbb{P}}, \mathbb{P}) = \tau(\hat{\mathbb{P}}) - \tau(\mathbb{P}) + \int \phi(o, \hat{\mathbb{P}}) d(\mathbb{P} - \hat{\mathbb{P}})(o)
$$

Since $\int \phi(o, \hat{\mathbb{P}}) d\hat{\mathbb{P}}(o) = 0$, this gives that

$$
R_2(\hat{\mathbb{P}}, \mathbb{P}) = \tau(\hat{\mathbb{P}}) - \tau(\mathbb{P}) + \int \phi(o, \hat{\mathbb{P}}) d\mathbb{P}(o)
$$

The rest follows from expanding the specific expressions:

$$
R_2(\hat{\mathbb{P}}, \mathbb{P}) = \tau(\hat{\mathbb{P}}) - \tau(\mathbb{P}) + \int \phi(o, \hat{\mathbb{P}}) d\mathbb{P}(o)
$$
$$
= -\tau(\mathbb{P}) + \frac{1}{\hat{\mathbb{P}}D} \int \Bigg( d(\hat{m}_1(x) - m_1(x)) + \frac{da}{\hat{\pi}_a(x)}(y - \hat{m}_1(x)) - \hat{w}(a, d, x)(y - \hat{m}_0(x)) \Bigg) d\mathbb{P}
$$
$$
= -\tau(\mathbb{P}) + \int (\hat{m}_1(x) - m_1(x)) f(x|d=1)dx + \int \frac{\pi_a(x)}{\hat{\pi}_a(x)}(\hat{m}_1(x) - m_1(x)) f(x|d=1)dx
$$
$$
- \int \frac{\hat{\pi}_d(x)}{\pi_d(x)}(m_0(x) - \hat{m}_0(x)) \frac{\pi_d(x)(1 - \pi_a(x)) + (1 - \pi_d(x))\hat{r}(x)}{\hat{\pi}_d(x)(1 - \hat{\pi}_d(x)) + (1 - \hat{\pi}_d(x))\hat{r}(x)} f(x|d=1)dx
$$
$$
= \int \Bigg( (m_0(x) - m_1(x)) + (\hat{m}_1(x) - \hat{m}_0(x)) + \frac{\pi_a(x)}{\hat{\pi}_a(x)}(m_1(x) - \hat{m}_1(x))
$$
$$
= - \frac{\hat{\pi}_d(x)}{\pi_d(x)}(m_0(x) - \hat{m}_0(x)) \frac{\pi_d(x)(1 - \pi_a(x)) + (1 - \pi_d(x))\hat{r}(x)}{\hat{\pi}_d(x)(1 - \hat{\pi}_d(x)) + (1 - \hat{\pi}_d(x))\hat{r}(x)} \Bigg) f(x|d=1)dx \quad \square
$$

**Lemma 4**: Assume that $||m_1(x) - \hat{m}_1(x)|| ||\pi_a(x) - \hat{\pi}_a(x)|| = o_{\mathbb{P}}(a_n)$, $||\hat{m}_0(x) - m_0(x)|| ||\pi_a(x) - \hat{\pi}_a(x)|| = o_{\mathbb{P}}(b_n)$, and $||\hat{m}_0(x) - m_0(x)|| ||\pi_d(x) - \hat{\pi}_d(x)|| = o_{\mathbb{P}}(c_n)$, where the norm is defined

with respect to the study distribution. Furthermore, assume that $0 < \epsilon \leqslant \hat{\pi}_a(x)$ and $\hat{r}(x) \leqslant M < \infty$ for all $x$ such that $f(x|d=1) > 0$. Then $R_2(\hat{\mathbb{P}}, \mathbb{P}) = o_{\mathbb{P}}(\max(a_n, b_n, c_n))$

**Proof**

By Lemma 3, we have that:

$$
R_2(\hat{\mathbb{P}}, \mathbb{P}) = \int \Bigg( (m_0(x) - m_1(x)) + (\hat{m}_1(x) - \hat{m}_0(x)) + \frac{\pi_a(x)}{\hat{\pi}_a(x)}(m_1(x) - \hat{m}_1(x)) \\
- \frac{\hat{\pi}_d(x)}{\pi_d(x)}(m_0(x) - \hat{m}_0(x)) \frac{\pi_d(x)(1 - \pi_a(x)) + (1 - \pi_d(x))\hat{r}(x)}{\hat{\pi}_d(x)(1 - \hat{\pi}_d(x)) + (1 - \hat{\pi}_d(x))\hat{r}(x)} \Bigg) f(x|d=1)dx.
$$

We first consider the expressions involving $m_1(x)$ and $\hat{m}_1(x)$.

$$
\int \Bigg( (\hat{m}_1(x) - m_1(x)) + \frac{\pi_a(x)}{\hat{\pi}_a(x)}(m_1(x) - \hat{m}_1(x)) \Bigg) f(x|d=1)dx
$$
$$
= \int \big( \hat{m}_1(x) - m_1(x) \big)\big( \pi_a(x) - \hat{\pi}_a(x) \big) \frac{1}{\hat{\pi}_a(x)} f(x|d=1)dx
$$
$$
\leqslant \int \big( \hat{m}_1(x) - m_1(x) \big)\big( \pi_a(x) - \hat{\pi}_a(x) \big) \frac{1}{\epsilon} f(x|d=1)dx
$$
$$
\leqslant ||m_1(x) - \hat{m}_1(x)|| ||\pi_a(x) - \hat{\pi}_a(x)|| \frac{1}{\epsilon}
$$
$$
= o_{\mathbb{P}}(a_n).
$$

The second to last inequality follows from Cauchy-Schwartz. Next, we consider the remainder expressions of $R_2(\hat{\mathbb{P}}, \mathbb{P})$. Letting $\lambda(x) = \hat{\pi}_d(x)\big( \pi_d(x)(1 - \pi_a(x)) + (1 - \pi_d(x))\hat{r}(x) \big)$ and $\hat{\lambda}(x) = \big( \pi_d(x)\hat{\pi}_d(x)(1 - \hat{\pi}_d(x)) + (1 - \hat{\pi}_d(x))\hat{r}(x) \big)$, we obtain the following:

$$\int \left( \big(m_0(x) - \hat{m}_0(x)\big) - \frac{\hat{\pi}_d(x)}{\pi_d(x)}(m_0(x) - \hat{m}_0(x)) \frac{\pi_d(x)(1 - \pi_a(x)) + (1 - \pi_d(x))\hat{r}(x)}{\hat{\pi}_d(x)(1 - \hat{\pi}_d(x)) + (1 - \hat{\pi}_d(x))\hat{r}(x)} \right) f(x|d=1)dx$$

$$= \int \big(m_0(x) - \hat{m}_0(x)\big)\big(1 - \frac{\lambda(x)}{\hat{\lambda}(x)}\big) f(x|d=1)dx$$

$$= \int \big(m_0(x) - \hat{m}_0(x)\big)\big(\hat{\lambda}(x) - \lambda(x)\big)\frac{1}{\hat{\lambda}(x)} f(x|d=1)dx$$

$$\leqslant \int \big(m_0(x) - \hat{m}_0(x)\big)\big(\hat{\lambda}(x) - \lambda(x)\big)\frac{1}{M} f(x|d=1)dx$$

$$\leqslant ||\hat{m}_0(x) - m_0(x)|| \hat{\lambda}(x) - \lambda(x)|| \frac{1}{M}$$

$$= ||\hat{m}_0(x) - m_0(x)|||| \hat{\pi}_d(x)\pi_d(x)\big(\pi_a(x) - \hat{\pi}_a(x)\big) + \big(\pi_d(x) - \hat{\pi}_d(x)\big)\hat{r}(x)|| \frac{1}{M}$$

$$\leqslant ||\hat{m}_0(x) - m_0(x)|||\pi_a(x) - \hat{\pi}_a(x)|||| + ||\hat{m}_0(x) - m_0(x)|||| \pi_d(x) - \hat{\pi}_d(x)|| \frac{1}{M}$$

$$= o_{\mathbb{P}}(b_n) + o_{\mathbb{P}}(c_n). \quad \square$$

**Theorem 2** Suppose that the regularity conditions in Lemma 2 hold for $\hat{\mathbb{P}}_{-k}$ for each $k$. Furthermore, assume that $||m_1(x) - \hat{m}_1(x)|||| \pi_a(x) - \hat{\pi}_a(x)|| = o_{\mathbb{P}}(1\backslash\sqrt{n})$, $||\hat{m}_0(x) - m_0(x)|||| \pi_a(x) - \hat{\pi}_a(x)|| = o_{\mathbb{P}}(1\backslash\sqrt{n})$, and $||\hat{m}_0(x) - m_0(x)|||| \pi_d(x) - \hat{\pi}_d(x)|| = o_{\mathbb{P}}(1\backslash\sqrt{n})$, where the norm is defined with respect to the study distribution and this holds for each cross-fit model. Furthermore, assume that $0 < \epsilon \leqslant \hat{\pi}_a(x)$ and $\hat{r}(x) \leqslant M < \infty$ for all $x$ such that $f(x|d=1) > 0$. Then:

$$\sqrt{n}(\hat{\tau}_{dr} - \tau) \xrightarrow{d} N(0, E[\phi(O, \mathbb{P})]).$$

### 1.13 *TMLE Fluctuation*

The estimator $\hat{\tau}_{aipw}$ was defined as the solution of an estimating equation so that $\mathbb{P}_n\{\phi(O, \hat{\mathbb{P}}, \hat{\tau}_{aipw}) = 0$. Alternatively, $\hat{\tau}_{tmle}$ is defined as $\tau(\hat{\mathbb{P}}^*) = \tau(\hat{\mathbb{P}}^*) + \mathbb{P}_n\{\phi(O, \hat{\mathbb{P}}^*)\}$, where $\hat{\mathbb{P}}^*)$ is a fluctuation of $\mathbb{P}$ so that $\mathbb{P}_n\{\phi(O, \hat{\mathbb{P}}^*)\} = 0$. Similar to Gruber and van der Laan (2010), we define a loss function $L(P) = (L_m(m_a), -log P_{X|D=1})$, where $P_{X|D=1}$ is the distribution of $X$ given $D = 1$, $m_a$ refers to $\{m_1(x), m_0(x)\}$, and

$$L_m(m_A)(O) = -Ylog(m_A(X)) - (1-Y)log(1-m_A(X))$$

It is shown in Gruber and van der Laan (2010) that this loss is minimized at $\mathbb{P}$. Furthermore, we define a parametric working model $\{P(\epsilon)\}$ with a 2-dimensional parameter $\epsilon = (\epsilon_1, \epsilon_2)$ for which $P(0) = \mathbb{P}$ and that $\frac{d}{d\epsilon}L(P(\epsilon))$ at $\epsilon = 0$ equals $\phi(O, \mathbb{P})$. We consider parametric working models given by

$$\text{logit}\big(m_a(\epsilon_1)\big) \equiv \text{logit}\big(m_a(x)\big) + \epsilon_1 h$$

and

$$P_{X|D=1}(\epsilon_2) = \left(1 + \epsilon_2 \frac{1}{q}\Big[Dm_1(X) - Dm_0(X) - D\tau(P)\Big]\right)P_{X|D=1}$$

where

$$h(D, A, X) \equiv \frac{1}{q}\left[\frac{DA}{\hat{\pi}_a(X)} - \frac{D(1-A)\hat{\pi}_d(X) + (1-D)(1-A)\hat{\pi}_d(X)\hat{r}(X)}{\hat{\pi}_d(X)(1-\hat{\pi}_a(X)) + (1-\hat{\pi}_d(X))\hat{r}(X)}\right]$$

It is easy to verify that, when $\epsilon = 0$, $P(\epsilon) = \mathbb{P}$ and $\frac{d}{d\epsilon}L(P(\epsilon)) = \phi(O, \mathbb{P})$. After one iteration, $\epsilon = 0$ so that $\mathbb{P}_n(O, \hat{\mathbb{P}}^*, \tau(\hat{\mathbb{P}}^*)) = 0$.

## 1.14 *Simulation Details*

In the simulations, three covariates, $X_1, X_2, X_3$, were generated independently. $X_1$ followed a Bernoulli(0.5) distribution while $X_2$ and $X_3$ both followed a Uniform(-1,1) distribution. In our simulation, we aimed to understand the properties of the estimators when $(X_1, X_2, X_3) \not\perp D$. Therefore, conditional upon these covariates, the study indicator $D$ was drawn from a Bernoulii distribution with mean $\pi_d(X_1, X_2, X_3)$, where

$$\pi_d(X_1, X_2, X_3) = \text{logit}^{-1}\bigg(f_{1a}(X_1) + f_{1b}(1 - X_1) + f_2(X_2) + f_3(X_3)\bigg).$$

We chose the basis functions in the following way:

$$f_{1a}(x) = \log\big(\tfrac{0.7}{0.3}\big)x$$

$$f_{2a}(x) = \log\big(\tfrac{0.3}{0.7}\big)x$$

$$f_2(x) = \log\bigg(2^{\frac{3}{7}}\frac{(1.5 - x^2)}{\frac{1}{6} + x^2}\bigg)$$

$$f_3(x) = \log\bigg(\frac{\frac{\sqrt{\pi}}{\text{erf}(1)}e^{-x^2}}{\frac{1.5 - e^{-x^2}}{3 - \sqrt{\pi}\text{erf}(1)}}\bigg).$$

This formulation allows us to directly control the relative ratio of the densities in the RCT and external control populations $(p(x_1, x_2, x_3|D = 1)/p(x_1, x_2, x_3|D = 0)$, where the ratio of the marginal densities is given by the fractions inside the logarithms) and therefore characterize differences in the distributions between the two populations. Under this simulation, when $D = 1$, $X_1 = 1$ more commonly and both $X_2$ and $X_3$ have greater density around 0. Because all of the covariates are bounded, the uniform overlap assumption on $\pi_d$ is satisfied.

For the study sample, one third of individuals were randomly assigned $A = 0$ (control) and two thirds were assigned $A = 1$ (treatment), independent of $(X_1, X_2, X_3)$. If $D = 0$ , then $A = 0$. The outcome $Y$ was generated as follows:

$$Y = 10 + X_1 - 2.5X_2^2 - 2.5X_3^2 + A\big(0.7 + 0.5X_1 + 0.2(1 - X_2^2)\big) + \epsilon.$$

Additionally, $\epsilon \sim N(0, 3)$ and is independent of any of the other variables.

To estimate the nuisance functions, two approaches were considered. When investigators had access to transformed versions of the variables (so that the logits and expected outcome were linear functions of the variables), a logistic regression model with main effects was

used for $\pi_d(x)$ and a linear regression model with main effects was used for $m_0(x)$ (and fit just to samples with $A = 0$). In these scenarios, the parameter estimates were consistent for their true values. When investigators did not have access to the transformed variables (Settings 2, 3, and 5), Random Forests were used via the randomForest package in R. For both nuisance models, forests with 500 trees and a *mtry* parameter (number of variables randomly sampled as candidates at each split) of 2 were fit. The maximum number of nodes for the tree was selected as 5 for the outcome model and 10 for the study propensity model. While the number of trees was chosen based on computational considerations, the maximum number of nodes and *mtry* parameter were chosen based on cross-validation from the first simulation iteration. We found that these hyperparameters tended to be fairly consistent in this simulation setting across different data set generations. All simulations were performed using R 4.1.0.

Confidence intervals for $\hat{\tau}_{om}$ and $\hat{\tau}_{ps}$ were constructed using the nonparametric bootstrap (percentile method). For $\hat{\tau}_{aipw}$ and $\hat{\tau}_{tmle}$, the estimators asymptotically follow a normal distribution with variance equal to the variance of the efficient influence function. Empirical estimates of this variance were calculated using the fitted nuisance functions and estimates of the average treatment effect.

## 2. SUNFISH Trial

SUNFISH (NCT02908685) (Mercuri et al., 2022) was a two-part multi-site randomized placebo-controlled trial designed to investigate the efficacy of risdiplam. In Part 2 of the study, 180 patients with Type 2 and non-ambulant Type 3 SMA were randomized 2:1 to receive risdiplam or placebo for 12 months. At the completion of 12 months, patients originally randomized to placebo were switched to risdiplam. While the SUNFISH trial did not incorporate external controls into its analysis, we use it as a hypothetical example of a trial that, were it conducted in the future, could benefit from the incorporation of external

controls. The external controls considered in this paper are from the placebo arm of a Phase 2 trial of olesoxime (NCT01302600). The trial is temporally and geographically similar, which could help justify the conditional mean exchangeability assumption. After restricting to complete cases with Type 2 or non-ambulant Type 3 SMA between the ages of 2-25, the RCT and external control samples contain 159 and 48 observations, respectively. While the original study performed a longitudinal analysis, for simplicity we restrict our attention to the change from baseline at 12 months. The primary endpoint of interest was the change in Motor Function Measure (MFM) at 12 months ($\Delta_{MFM}$). Several different scales exist for MFM. In order to ensure the comparability of MFM scores between the placebo group of the external control and the patients in SUNFISH, the change in baseline in the MFM total score is derived from MFM32 for all patients aged 6 years or above and MFM20 for all patients aged less than 6 years.

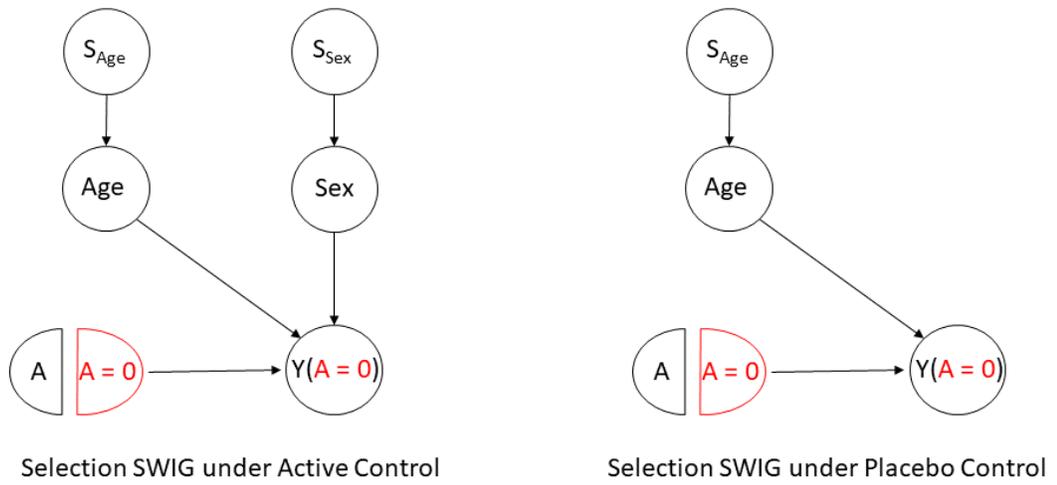

Selection SWIG under Active Control          Selection SWIG under Placebo Control

**Figure 1.**



\end{document}